\RequirePackage{ifpdf}
\ifpdf 
\documentclass[pdftex]{sigma}
\else
\documentclass{sigma}
\fi

\numberwithin{equation}{section}
\numberwithin{theorem}{section}
\numberwithin{proposition}{section}
\numberwithin{lemma}{section}
\numberwithin{corollary}{section}
\numberwithin{definition}{section}
\numberwithin{example}{section}
\numberwithin{remark}{section}
\numberwithin{note}{section}

\def\aa{\alpha}
\def\dl{\delta}
\def\ep{\varepsilon}
\def\ka{\kappa}
\def\sig{\sigma}
\def\Lam{\Lambda}
\def\bs{\backslash}
\def\wt{\widetilde}

\def\wa{\widetilde{a}}
\def\ch{\check}
\def\disp{\displaystyle}
\def\hf{\frac{1}{2}}

\def\I{\sqrt{-1}}
\def\Ra{\Rightarrow}
\def\lra{\leftrightarrow}
\def\Lra{\Leftrightarrow}
\newcommand{\br}[1]{{\langle{#1}\rangle}}
\newcommand{\BZ}{\mathbb Z}
\newcommand{\BC}{\mathbb C}
\newcommand{\BP}{\mathbb P}
\newcommand{\Da}{\mathfrak{a}}
\newcommand{\Db}{\mathfrak{b}}
\newcommand{\Dc}{\mathfrak{c}}
\newcommand{\CA}{{\cal A}}
\newcommand{\CB}{{\cal B}}
\newcommand{\CC}{{\cal C}}
\newcommand{\CE}{{\cal E}}
\newcommand{\CG}{{\cal G}}
\newcommand{\CN}{{\cal N}}
\newcommand{\CL}{{\cal L}}
\newcommand{\DS}{\mathfrak{S}}
\newcommand{\ol}{\overline}
\newcommand{\ul}{\underline}
\newcommand{\e}{{\rm e}}

\begin{document}

\allowdisplaybreaks

\renewcommand{\thefootnote}{$\star$}

\renewcommand{\PaperNumber}{035}

\FirstPageHeading

\ShortArticleName{Hypergeometric $\tau$-Functions of the $q$-Painlev\'e System of Type $E_7^{(1)}$}

\ArticleName{Hypergeometric $\boldsymbol{\tau}$-Functions\\ of the $\boldsymbol{q}$-Painlev\'e System of Type $\boldsymbol{E_7^{(1)}}$\footnote{This paper is a contribution to the Proceedings of the Workshop ``Elliptic Integrable Systems, Isomonodromy Problems, and Hypergeometric Functions'' (July 21--25, 2008, MPIM, Bonn, Germany). The full collection
is available at
\href{http://www.emis.de/journals/SIGMA/Elliptic-Integrable-Systems.html}{http://www.emis.de/journals/SIGMA/Elliptic-Integrable-Systems.html}}}

\Author{Tetsu MASUDA}

\AuthorNameForHeading{T. Masuda}

\Address{Department of Physics and Mathematics,
Aoyama Gakuin University,\\
5-10-1 Fuchinobe, Sagamihara, Kanagawa, 229-8558, Japan}

\Email{\href{mailto:masuda@gem.aoyama.ac.jp}{masuda@gem.aoyama.ac.jp}}

\ArticleDates{Received November 27, 2008, in f\/inal form March 10,
2009; Published online March 24, 2009}

\Abstract{We present the $\tau$-functions for the hypergeometric solutions to the $q$-Painlev\'e system of type $E_7^{(1)}$ in a determinant formula whose entries are given by the basic hypergeometric function ${}_8W_7$. By using the $W(D_5)$ symmetry of the function ${}_8W_7$, we construct a~set of twelve solutions and describe the action of $\wt{W}(D_6^{(1)})$ on the set.}


\Classification{33D15; 33D05; 33D60; 33E17}

\renewcommand{\thefootnote}{\arabic{footnote}}
\setcounter{footnote}{0}

\section{Introduction}

A natural framework for discrete Painlev\'e equations by means of the geometry of rational surfaces has been proposed by Sakai~\cite{Sakai2}. Each equation is def\/ined by the group of Cremona transformations on a family of surfaces obtained by blowing-up at nine points on $\BP^2$. According to the types of rational surfaces, the discrete Painlev\'e equations are classif\/ied in terms of af\/f\/ine root systems. Also, their symmetries are described by means of af\/f\/ine Weyl groups, the lattice part of which gives rise to dif\/ference equations. For instance, the $q$-Painlev\'e system of type $E_7^{(1)}$, which is the main object of this paper, is a discrete dynamical system def\/ined on a family of rational surfaces parameterized by nine-point conf\/igurations on $\BP^2$ such that six points among them are on a conic and other three are on a line~\cite{Sakai2}. An explicit expression for the system of $q$-dif\/ference equations is given by~\cite{RGTT}
\begin{gather}
\dfrac{(f\ol{g}-\ol{t}t)(fg-t^2)}
      {(f\ol{g}-1)(fg-1)}
=\dfrac{(f-b_1t)(f-b_2t)(f-b_3t)(f-b_4 t)}
       {(f-b_5)(f-b_6)(f-b_7)(f-b_8)},\nonumber\\
\dfrac{(fg-t^2)(\ul{f}g-t\ul{t})}
      {(fg-1)(\ul{f}g-1)}
=\dfrac{\left(g-\frac{t}{b_1}\right)\left(g-\frac{t}{b_2}\right)
        \left(g-\frac{t}{b_3}\right)\left(g-\frac{t}{b_4}\right)}
       {\left(g-\frac{1}{b_5}\right)\left(g-\frac{1}{b_6}\right)
        \left(g-\frac{1}{b_7}\right)\left(g-\frac{1}{b_8}\right)},
  \label{eq:E7}
\end{gather}
where $t$ is the independent variable and the time evolution of the dependent variables is given by $\ol{g}=g(qt)$ and $\ul{f}=f(t/q)$. The parameters $b_i$ $(i=1,2,\ldots,8$) satisfy $b_1b_2b_3b_4=q$ and $b_5b_6b_7b_8=1$.

Similarly to the Painlev\'e dif\/ferential equations, the discrete Painlev\'e equations admit particular solutions expressible in terms of various hypergeometric functions. Regarding the $q$-dif\/ference Painlev\'e equations, the hypergeometric solutions to those equations have been constructed by means of a geometric approach and direct linearization of the $q$-dif\/ference Riccati equations~\cite{KMNOY2,KMNOY3}. In particular, the Riccati solution to the system of $q$-dif\/ference equations (\ref{eq:E7}) is expressed in terms of the $q$-hypergeometric series
\begin{gather}
{}_8W_7(a_0;a_1,\ldots,a_5;q,z) ={}_8\varphi_7
\left(
\begin{array}{c}
a_0,qa_0^{1/2},-qa_0^{1/2},a_1\cdots,a_5\\[1mm]
a_0^{1/2},-a_0^{1/2},qa_0/a_1,\cdots,qa_0/a_5
\end{array}
;q,z
\right)\nonumber\\
 \qquad{} =\sum_{k=0}^{\infty}
\dfrac{(1-a_0q^{2k})}{(1-a_0)}\,
\dfrac{(a_0;q)_k}{(q;q)_k}
\prod_{i=1}^5\dfrac{(a_i;q)_k}{(qa_0/a_i;q)_k}\, z^k,
\qquad z=\dfrac{q^2a_0^2}{a_1a_2a_3a_4a_5},
   \label{8W7}
\end{gather}
where $(a;q)_k=\prod\limits_{i=0}^{k-1}(1-aq^i)$.

The purposes of this paper are to propose a formulation for the $q$-Painlev\'e system of type~$E_7^{(1)}$ by means of the lattice $\tau$-functions and to completely determine the $\tau$-functions for the hypergeometric solutions (hypergeometric $\tau$-functions for short) of the system.

This paper is organized as follows. In Section~\ref{qP_E7}, we give a formulation for the $q$-Painlv\'e system of type $E_7^{(1)}$ in terms of the lattice $\tau$-functions. Section~\ref{pre} is devoted to a preparation for constructing the hypergeometric $\tau$-functions. We decompose the lattice, each of whose elements indicates the $\tau$-function, into a family of six-dimensional lattices.

In Sections \ref{M0}--\ref{det}, we construct the hypergeometric $\tau$-functions. We f\/ind that a $q$-analogue of the double gamma function appears as a normalization factor of the hypergeometric $\tau$-functions in Section~\ref{M0}. In Section \ref{M1}, we f\/ind that a class of bilinear equations for the lattice $\tau$-functions yields the contiguity relations for the $q$-hypergeometric function ${}_8W_7$. As is well-known, the $q$-hypergeometric function ${}_8W_7$ possesses the $W(D_5)$-symmetry~\cite{LJ1}. From that, we can construct a set of twelve solutions corresponding to the coset $W(D_6)/W(D_5)$, and describe the action of $\wt{W}\big(D_6^{(1)}\big)$ on the set of solutions.

One of the important features of the hypergeometric solutions to the continuous and discrete Painlev\'e equations is that they can be expressed in terms of Wronskians or Casorati determinants~\cite{HKW,KNY,KK,HK,Sakai1}. In Section \ref{det}, we show that the hypergeometric $\tau$-functions of the $q$-Painlev\'e system of type $E_7^{(1)}$ are expressed by ``two-directional Casorati determinants''. As a~consequence, we get an explicit expression for the hypergeometric solutions to the $q$-dif\/ference Painlev\'e equation~(\ref{eq:E7}), which is proposed in Corollary~\ref{sol_diff_eq}.

\section[The $q$-Painlev\'e system of type $E_7^{(1)}$]{The $\boldsymbol{q}$-Painlev\'e system of type $\boldsymbol{E_7^{(1)}}$}\label{qP_E7}

\subsection[The discrete Painlev\'e system of type $E_8^{(1)}$]{The discrete Painlev\'e system of type $\boldsymbol{E_8^{(1)}}$}

At f\/irst, we give a brief review of the formulation for the discrete Painlev\'e system of type $E_8^{(1)}$ in terms of the lattice~$\tau$-functions~\cite{KMNOY1,KMNOY4}.

Let $\CL=\mathop{\oplus}\limits_{i=0}^9 \BZ \e_i$ be a lattice with a basis $\{\e_0,\e_1,\ldots,\e_9\}$, and def\/ine a symmetric bilinear form $\br{~,~}: \CL\times\CL\to\BZ$ by
\begin{gather*}
\br{\e_0,\e_0}=-1,\qquad\br{\e_i,\e_i}=1\quad (i=1,2,\ldots,9),\qquad
\br{\e_i,\e_j}=0\quad (i,j=0,1,\ldots,9;\ i\ne j).
\end{gather*}
Consider the af\/f\/ine Weyl group $W(E_8^{(1)})=\br{s_0,s_1,\ldots,s_8}$ associated with the Dynkin diagram
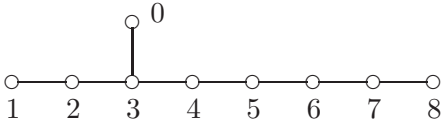
\[
\setlength{\unitlength}{0.8mm}
\begin{picture}(90,15)(0,-5)
\put(-1,-6){1}
\put(9,-6){2}
\put(19,-6){3}
\put(29,-6){4}
\put(39,-6){5}
\put(49,-6){6}
\put(59,-6){7}
\put(69,-6){8}
\put(23,10){0}
\put(0,0){\circle{2}}
\put(10,0){\circle{2}}
\put(20,0){\circle{2}}
\put(30,0){\circle{2}}
\put(40,0){\circle{2}}
\put(50,0){\circle{2}}
\put(60,0){\circle{2}}
\put(70,0){\circle{2}}
\put(20,10){\circle{2}}
\put(1,0){\line(1,0){8}}
\put(11,0){\line(1,0){8}}
\put(21,0){\line(1,0){8}}
\put(31,0){\line(1,0){8}}
\put(41,0){\line(1,0){8}}
\put(51,0){\line(1,0){8}}
\put(61,0){\line(1,0){8}}
\put(20,1){\line(0,1){8}}
\end{picture}
\]
The lattice $\CL$ admits a natural linear action of $W(E_8^{(1)})$ def\/ined by $s_i\,.\, \Lam=\Lam-\br{h_i,\Lam} h_i$ for $\Lam\in\CL$, where $h_i$ $(i=0,1,\ldots,8)$ are the simple coroots def\/ined by $h_0=\e_0-\e_1-\e_2-\e_3$ and $h_i=\e_i-\e_{i+1}$ $(i=1,\ldots,8)$. The canonical central element $c=3\e_0-\e_1-\cdots-\e_9$ is orthogonal to all the simple coroots $h_i$, and hence $W(E_8^{(1)})$-invariant.

The parameter space for the discrete Painlev\'e system of type $E_8^{(1)}$ is the ten-dimensional vector space $\mathop{\oplus}\limits_{i=0}^9 \BC \e_i$, whose coordinates are denoted by $\ep_i=\br{\e_i,\cdot}$ $(i=0,1,\ldots,9)$. The root lattice $Q(E_8^{(1)})=\mathop{\oplus}\limits_{i=0}^8 \BZ \aa_i$ is generated by the simple roots $\aa_0=\ep_0-\ep_1-\ep_2-\ep_3$ and $\aa_i=\ep_i-\ep_{i+1}$ $(i=1,\ldots,8)$. The af\/f\/ine Weyl group $W(E_8^{(1)})$ acts on the coordinate function $\ep_i$ in a similar way to on the basis $\e_i$. The $W(E_8^{(1)})$-invariant element corresponding to $c$ is given by $\dl=\br{c,\cdot}=3\ep_0-\ep_1-\cdots-\ep_9$, which is called the null root and plays the role of the scaling constant for dif\/ference equations in the context of the discrete Painlev\'e equations. For simplicity, we denote the ref\/lection $s_{\aa}$ with respect to the root $\aa=\ep_{ij}=\ep_i-\ep_j$ or $\aa=\ep_{ijk}=\ep_0-\ep_i-\ep_j-\ep_k$ for $i,j,k\in\{1,2,\ldots,9\}$ by $s_{ij}$ or $s_{ijk}$, respectively. Also, we often use the notation $\e_{ij}=\e_i-\e_j$ and $\e_{ijk}=\e_0-\e_i-\e_j-\e_k$.

For each $\aa\in Q\big(E_8^{(1)}\big)$, the action of the translation operator $T_{\aa}\in W\big(E_8^{(1)}\big)$ is given by~\cite{Kac}
\begin{equation}
T_{\aa}(\Lam)=\Lam+\br{c,\Lam}\,h
-\left(\dfrac{1}{2}\br{h,h}\br{c,\Lam}+\br{h,\Lam}\right)c
\qquad (\Lam\in\CL) \label{def_trans}
\end{equation}
by using the element $h\in\CL$ such that $\aa=\br{h,\cdot}$. Note that we have $T_{\aa}T_{\beta}=T_{\beta}T_{\aa}$ and $wT_{\aa}w^{-1}=T_{w.\aa}$ for any $w\in W\big(E_8^{(1)}\big)$. When $\aa=\ep_{ij}$ or $\ep_{ijk}$, we also denote the translation $T_{\aa}$ simply by $T_{ij}$ or $T_{ijk}$, respectively. They can be expressed by
\begin{alignat*}{3}
&T_{ij}=s_{il_1l_2}s_{il_3l_4}s_{l_5l_6l_7}s_{il_3l_4}s_{il_1l_2}s_{ij},\qquad &&
\{i,j,l_1,\ldots,l_7\}=\{1,2,\ldots,9\},& \\
& T_{ijk}=s_{l_1l_2l_3}s_{l_4l_5l_6}s_{l_1l_2l_3}s_{ijk},\qquad &&
\{i,j,k,l_1,\ldots,l_6\}=\{1,2,\ldots,9\}.&
\end{alignat*}

Let us introduce a family of dependent variables $\tau_{\Lam}=\tau_{\Lam}(\ep)$, $\ep=(\ep_0,\ldots,\ep_9)$, indexed by $\Lam\in M$, where $M$ is the $W(E_8^{(1)})$-orbit def\/ined by
\[
M=W\big(E_8^{(1)}\big)\,.\, \e_1
=\{\Lam\in\CL\,|\,\br{c,\Lam}=-1,\br{\Lam,\Lam}=1\}\subset\CL.
\]
The action of $W\big(E_8^{(1)}\big)$ on the lattice $\tau$-functions $\tau_{\Lam}$ is def\/ined by $w(\tau_{\Lam})=\tau_{w.\Lam}$ for any $w\in W\big(E_8^{(1)}\big)$. The discrete Painlev\'e system of type $E_8^{(1)}$ is equivalent to the overdetermined system def\/ined by the bilinear equations
\[
 [\ep_{jk}][\ep_{jkl}]\tau_{\e_i}\tau_{\e_0-\e_i-\e_l}
+[\ep_{ki}][\ep_{kil}]\tau_{\e_j}\tau_{\e_0-\e_j-\e_l}
+[\ep_{ij}][\ep_{ijl}]\tau_{\e_k}\tau_{\e_0-\e_k-\e_l}=0
\]
for any mutually distinct indices $i,j,k,l\in\{1,2,\ldots,9\}$, as well as their $W\big(E_8^{(1)}\big)$-transforms
\[
[w(\ep_{jk})][w(\ep_{jkl})]\tau_{w.\e_i}\tau_{w.(\e_0-\e_i-\e_l)}
+(i,j,k)\mbox{-cyclic}=0
\]
for any $w\in W\big(E_8^{(1)}\big)$. Here, $[x]$ is a nonzero odd holomorphic function on $\BC$ satisfying the Riemann relation
\[
[x+y][x-y][u+v][u-v]=[x+u][x-u][y+v][y-v]-[x+v][x-v][y+u][y-u]
\]
for any $x,y,u,v\in\BC$. There are three classes of such functions; elliptic, trigonometric and rational. These three cases correspond to the three types of dif\/ference equations, namely, elliptic dif\/ference, $q$-dif\/ference and ordinal dif\/ference, respectively. The lattice part of $W\big(E_8^{(1)}\big)$ gives rise to the dif\/ference Painlev\'e equation.

\subsection[The $q$-Painlev\'e system of type $E_7^{(1)}$]{The $\boldsymbol{q}$-Painlev\'e system of type $\boldsymbol{E_7^{(1)}}$}

Let us propose a formulation for the $q$-Painlev\'e system of type $E_7^{(1)}$ by means of the lattice $\tau$-functions, using by the notation introduced in the previous subsection. A derivation of the formulation is discussed in Appendices.

The $q$-Painlev\'e system of type $E_7^{(1)}$ is a discrete dynamical system def\/ined on a family of rational surfaces parameterized by nine-point conf\/igurations on $\BP^2$ such that six points among them are on a conic $C$ and other three are on a line $L$~\cite{Sakai2}. Here, we set $p_1,p_2,p_3,p_4,p_5,p_6\in C$ and $p_7,p_8,p_9\in L$. In what follows, the symbols $C$ and $L$ also mean the index sets $C=\{1,2,3,4,5,6\}$ and $L=\{7,8,9\}$, respectively. And we often use $i,j,k,\dots$ and $r,s$ as the elements of $C$ and $L$, respectively. In this setting, the symmetric groups $\DS_6^C=\br{s_{12},\ldots,s_{56}}$ and $\DS_3^L=\br{s_{78},s_{89}}$ naturally act on the conf\/iguration space as the permutation of the points on~$C$ and~$L$, respectively. Also, the standard Cremona transformation with respect to $(p_1,p_2,p_7)$ is well-def\/ined as a birational action on the space. They generate the af\/f\/ine Weyl group $W(E_7^{(1)})=\br{s_{12},s_{23},s_{34},s_{45},s_{56},s_{78},s_{89},s_{127}}$. The associated Dynkin diagram and its automorphism are realized by
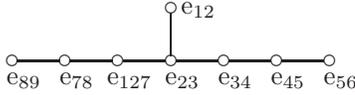
\[
\begin{picture}(140,25)(0,10)
\put(0,15){\circle{4}}\put(2,15){\line(1,0){16}}
\put(20,15){\circle{4}}\put(22,15){\line(1,0){16}}
\put(40,15){\circle{4}}\put(42,15){\line(1,0){16}}
\put(60,15){\circle{4}}\put(62,15){\line(1,0){16}}
\put(80,15){\circle{4}}\put(82,15){\line(1,0){16}}
\put(100,15){\circle{4}}\put(102,15){\line(1,0){16}}
\put(120,15){\circle{4}}
\put(60,35){\circle{4}}\put(60,17){\line(0,1){16}}
\put(-2,6){\small$\e_{89}$}
\put(18,6){\small$\e_{78}$}
\put(36,6){\small$\e_{127}$}
\put(58,6){\small$\e_{23}$}
\put(78,6){\small$\e_{34}$}
\put(98,6){\small$\e_{45}$}
\put(118,6){\small$\e_{56}$}
\put(64,33){\small$\e_{12}$}
\end{picture}
\]
and $\pi=s_{123}s_{47}s_{58}s_{69}$, respectively. Thus we f\/ind that the extended af\/f\/ine Weyl group $\wt{W}\big(E_7^{(1)}\big)$ $=\br{s_{12},s_{23},s_{34},s_{45},s_{56},s_{78},s_{89},s_{127},\pi}$ acts on the conf\/iguration space.

The lattice $\tau$-functions $\tau_{\Lam}=\tau_{\Lam}(\ep)$ for the $q$-Painlev\'e system of type $E_7^{(1)}$ are indexed by
\[
\Lam\in M^{E_7}=\wt{W}\big(E_7^{(1)}\big)\, .\, \e_1=M^C\coprod M^L,
\] where
\begin{gather*}
M^C=\{\Lam\in M\,|\,\br{\e_{789},\Lam}= 0\}=W\big(E_7^{(1)}\big)\,.\,\e_1,\\
M^L=\{\Lam\in M\,|\,\br{\e_{789},\Lam}=-1\}=W\big(E_7^{(1)}\big)\,.\,\e_7.
\end{gather*}
The action of $\wt{W}(E_7^{(1)})$ on the lattice $\tau$-functions $\tau_{\Lam}$ is def\/ined by $w(\tau_{\Lam})=\tau_{w.\Lam}$ for any $w\in\wt{W}\big(E_7^{(1)}\big)$. The $q$-Painlev\'e system of type $E_7^{(1)}$ is equivalent to the overdetermined system def\/ined by the bilinear equations
\begin{gather}
[\ep_{rs}]\tau_{\e_j}^C\tau_{\e_0-\e_i-\e_j}^L
=[\ep_{ijs}]\tau_{\e_r}^L\tau_{\e_0-\e_i-\e_r}^C
-[\ep_{ijr}]\tau_{\e_s}^L\tau_{\e_0-\e_i-\e_s}^C,\nonumber\\
[\ep_{jk}]\tau_{\e_r}^L\tau_{\e_0-\e_i-\e_r}^C
=[\ep_{ikr}]\tau_{\e_j}^C\tau_{\e_0-\e_i-\e_j}^L
-[\ep_{ijr}]\tau_{\e_k}^C\tau_{\e_0-\e_i-\e_k}^L,
  \label{bi_1}
\\
[\ep_{ij}][\ep_{ijr}]\tau_{\e_k}^C\tau_{\e_0-\e_k-\e_r}^C
+(i,j,k)\mbox{-cyclic}=0,\nonumber \\
[\ep_{ij}][\ep_{kl}]\tau_{\e_0-\e_i-\e_j}^L\tau_{\e_0-\e_k-\e_l}^L
+(i,j,k)\mbox{-cyclic}=0,
\label{bi_2}
\\
\tau_{\e_i}^C\tau_{\e_0-\e_i-\e_9}^C-\tau_{\e_j}^C\tau_{\e_0-\e_j-\e_9}^C
+[\ep_{ij}][\ep_{ij9}]\,d_L\tau_{\e_7}^L\tau_{\e_8}^L=0,\nonumber\\
\tau_{\e_0-\e_1-\e_4}^L\tau_{\e_0-\e_2-\e_3}^L-
\tau_{\e_0-\e_1-\e_3}^L\tau_{\e_0-\e_2-\e_4}^L
+[\ep_{12}][\ep_{34}]\,d_C\tau_{\e_5}^C\tau_{\e_6}^C=0
   \label{bi_3}
\end{gather}
for mutually distinct indices $i,j,k,l\in C$ and $r,s\in L$, as well as their $\wt{W}(E_7^{(1)})$-transforms. The superscript $C$ (resp.~$L$) denotes that the $\tau$-function is indexed by $\Lam\in M^C$ (resp.\ \mbox{$\Lam\in M^L$}), and we leave it out when it is unnecessarily. It is possible to f\/ix the function $[x]$ as $[x]=e(\hf x)-e(-\hf x)$, $e(x)=e^{\pi\I x}$, without loss of generality. The factors $d_L$ and $d_C$ in (\ref{bi_3}) correspond to the irreducible components of the anti-canonical devisor ${\cal D}_L=\e_0-\e_7-\e_8-\e_9$ and ${\cal D}_C=2\e_0-\e_1-\cdots-\e_6$, respectively. These factors are $W\big(E_7^{(1)}\big)$-invariant and the action of $\pi$ is given by $\pi:d_L\lra d_C$.

The translation operators with respect to the root vectors $\ep_{ij},\ep_{rs},\ep_{ijr}\in Q\big(E_7^{(1)}\big)$ are denoted by $T_{ij}$, $T_{rs}$ and $T_{ijr}$, respectively. Also, there exist f\/ifty six translation operators that move a~lattice point $\Lam\in M^{E_7}$ to its nearest ones. Let us denote such operators by $\wt{T}_{ir}$, $\wt{T}_{ijk}$ and $\wt{T}_{irs}$ according to the action on $Q\big(E_7^{(1)}\big)$;
\begin{alignat*}{4}
& \wt{T}_{17} : \quad&&
 \ep_{78}\mapsto\ep_{78}+\dl,\qquad &&\ep_{12}\mapsto\ep_{12}-\dl,& \\
& \wt{T}_{123}:\quad && \ep_{127}\mapsto\ep_{127}-\dl,\qquad &&\ep_{34}\mapsto\ep_{34}+\dl,&
\end{alignat*}
for instance. We f\/ind that these operators can be realized as $\wt{T}_{ir}=T_{ir}s_{789}$, $\wt{T}_{ijk}=s_{789}T_{ijk}$ and $\wt{T}_{irs}=T_{irs}s_{789}$, respectively, in terms of the Weyl group $W\big(E_8^{(1)}\big)$. Then, from the formula (\ref{def_trans}), the action on a lattice point can be calculated as $\wt{T}_{17}(\e_9)=T_{17}(\e_0-\e_7-\e_8)=\e_0-\e_1-\e_8$, for example. Note that we have the relations such as $\wt{T}_{19}\wt{T}_{178}=1$ and $\wt{T}_{123}\wt{T}_{456}=1$. The translation operators with respect to the root vectors can be expressed by $T_{ij}=\wt{T}_{ir}\wt{T}^{-1}_{jr}$, $T_{rs}=\wt{T}_{ir}^{-1}\wt{T}_{is}$ and $T_{ijr}=\wt{T}_{ijk}\wt{T}_{kr}$.

\begin{proposition}
If the lattice $\tau$-functions $\tau_{\Lam}\,(\Lam\in M^{E_7})$ satisfy the bilinear equations \eqref{bi_1} and their $\wt{W}\big(E_7^{(1)}\big)$-transforms, then they also satisfy \eqref{bi_2} and their $\wt{W}\big(E_7^{(1)}\big)$-transforms.
\end{proposition}

This is easily verif\/ied by a direct calculation. From this proposition, we see that it is not necessary to consider the bilinear equations~(\ref{bi_2}) for constructing a solution to the $q$-Painlev\'e system of type $E_7^{(1)}$. However, as we will see Section~\ref{det}, we use the bilinear equations of type~(\ref{bi_2}) in order to get a nicer determinant formula for the hypergeometric $\tau$-functions. Then, we will treat all types of bilinear equations below, although the discussion becomes technically complicated as a consequence.

Let us introduce the dependent variables $f$ and $g$ by
\begin{gather*}
f=e\left(\tfrac{1}{8}\aa_l-\tfrac{1}{8}\aa_r+\tfrac{1}{4}\ep_{12}+\tfrac{1}{8}\dl\right)
\dfrac{e( \tfrac{1}{4}\ep_{13})\tau_{\e_1}\tau_{\e_0-\e_1-\e_2}
      -e(-\frac{1}{4}\ep_{13})\tau_{\e_3}\tau_{\e_0-\e_2-\e_3}}
      {e( \frac{1}{4}\ep_{13})\tau_{\e_3}\tau_{\e_0-\e_2-\e_3}
      -e(-\frac{1}{4}\ep_{13})\tau_{\e_1}\tau_{\e_0-\e_1-\e_2}},\\
g=e\left(\tfrac{1}{8}\aa_r-\tfrac{1}{8}\aa_l+\tfrac{1}{4}\ep_{12}-\tfrac{1}{8}\dl\right)
\dfrac{e( \frac{1}{4}\ep_{23})\tau_{\e_3}\tau_{\e_0-\e_1-\e_3}
      -e(-\frac{1}{4}\ep_{23})\tau_{\e_2}\tau_{\e_0-\e_1-\e_2}}
      {e( \frac{1}{4}\ep_{23})\tau_{\e_2}\tau_{\e_0-\e_1-\e_2}
      -e(-\frac{1}{4}\ep_{23})\tau_{\e_3}\tau_{\e_0-\e_1-\e_3}}
\end{gather*}
with $\aa_l=3\ep_{127}+2\ep_{78}+\ep_{89}$ and $\aa_r=3\ep_{34}+2\ep_{45}+\ep_{56}$. Then, one get the explicit expression for the $q$-dif\/ference equations (\ref{eq:E7}), a derivation of which is discussed in Appendix~\ref{derivation}.

\section{A family of six-dimensional lattices and the bilinear equations\label{pre}}

As a preparation for constructing the hypergeometric $\tau$-functions, we decompose the lattice $M^{E_7}$ into a family of six-dimensional lattices according to the value of the symmetric bilinear form with the coroot vector $\e_{89}=\e_8-\e_9$;
\[
M^{E_7}=\coprod_{n\in\BZ}M_n,\qquad
M_n=\big\{\Lam\in M^{E_7}\,|\,\br{\Lam,\e_{89}}=n\big\}.
\]
Parallel to this decomposition, let us consider the orthogonal complement of $\ep_{89}$ in the root lattice $Q\big(E_7^{(1)}\big)$. Then we get the root lattice $Q\big(D_6^{(1)}\big)$ corresponding to the Dynkin diagram
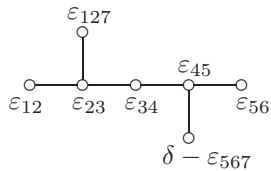
\[
\begin{picture}(140,50)(0,-25)
\put(20,0){\circle{4}}\put(22,0){\line(1,0){16}}
\put(40,0){\circle{4}}\put(42,0){\line(1,0){16}}
\put(60,0){\circle{4}}\put(62,0){\line(1,0){16}}
\put(80,0){\circle{4}}\put(82,0){\line(1,0){16}}
\put(100,0){\circle{4}}
\put(40,20){\circle{4}}\put(40,2){\line(0,1){16}}
\put(80,-20){\circle{4}}\put(80,-18){\line(0,1){16}}
\put(12,-9){\small$\ep_{12}$}
\put(36,-9){\small$\ep_{23}$}
\put(56,-9){\small$\ep_{34}$}
\put(76,5){\small$\ep_{45}$}
\put(98,-9){\small$\ep_{56}$}
\put(34,25){\small$\ep_{127}$}
\put(70,-30){\small$\dl-\ep_{567}$}
\end{picture}
\]
Since we have $\ep_{127}+\ep_{12}+2\ep_{23}+2\ep_{34}+2\ep_{45}+\ep_{56}+(\dl-\ep_{567})=\dl$, the same $\dl$ denotes the null root of $Q\big(D_6^{(1)}\big)$. The corresponding simple ref\/lections generate the af\/f\/ine Weyl group $W\big(D_6^{(1)}\big)=\br{s_{127},s_{12},s_{23},s_{34},s_{45},s_{56},s_{\dl-\ep_{567}}}$. Note that the f\/inite Weyl group $W(D_6)=\br{s_{127},s_{12},s_{23},s_{34},s_{45},s_{56}}$ includes the symmetric group $\DS_6=\br{s_{12},s_{23},s_{34},s_{45},s_{56}}$ as a subgroup. In this realization, an automorphism of the above Dynkin diagram can be expressed by $\rho=\pi s_{157}s_{168}s_{24}s_{26}s_{35}s_{79}$ whose action on the simple roots of $Q\big(D_6^{(1)}\big)$ is given by
\[
\rho :\quad
\ep_{12}\lra\dl-\ep_{567},\qquad
\ep_{127}\lra\ep_{56},\qquad
\ep_{23}\lra\ep_{45}.
\]
The extended af\/f\/ine Weyl group $\wt{W}(D_6^{(1)})=\br{s_{127},s_{12},s_{23},s_{34},s_{45},s_{56},s_{\dl-\ep_{567}},\rho}$ acts transitively on each $M_n$. Regarding the translation operators, we have $\wt{T}_{i7},\,\wt{T}_{ijk}\in\wt{W}\big(D_6^{(1)}\big)$ for $i,j,k\in C=\{1,2,\ldots,6\}$, which can be expressed in the form $\wt{T}_{\aa}=\rho w$, $w\in W\big(D_6^{(1)}\big)$.

According to the location of the lattice $\tau$-functions, one can classify the bilinear equa\-tions~(\ref{bi_1}) into the following four types:
\begin{alignat*}{3}
& \mbox{(A)}_n:\quad &&
\mbox{Two on each of $M_{n-1}$, $M_n$ and $M_{n+1}$, respectively}. & \\
& \mbox{(B)}_n:&&
\mbox{Four on $M_n$, and one on $M_{n+1}$ and $M_{n-1}$, respectively}. & \\
& \mbox{(C)}_n:&&\mbox{Three on $M_{n+1}$ and $M_n$, respectively}. &  \\
& \mbox{(D)}_n:&& \mbox{Six on $M_n$}. &
\end{alignat*}
The bilinear equations of type (C)$_n$ are further classif\/ied into two types. The f\/irst one is that all of three $\tau$-functions on $M_{n+1}$ belong to $M^C$ (or $M^L$), which is denoted by (C)$^{\rm r}_n$. The second is that one of three $\tau$-functions on $M_{n+1}$ belongs to $M^C$ (or $M^L$), denoted by (C)$^{\rm i}_n$. Typical bilinear equations are given by
\begin{alignat}{3}
& \mbox{(A)}_0\quad &&
[\ep_{89}]\tau_{\e_j}\tau_{\e_0-\e_i-\e_j}
=[\ep_{ij9}]\tau_{\e_8}\tau_{\e_0-\e_i-\e_8}
-[\ep_{ij8}]\tau_{\e_9}\tau_{\e_0-\e_i-\e_9},&\nonumber\\
& \mbox{(B)}_0 &&
[\ep_{78}]\tau_{\e_j}\tau_{\e_0-\e_i-\e_j}
=[\ep_{ij8}]\tau_{\e_7}\tau_{\e_0-\e_i-\e_7}
-[\ep_{ij7}]\tau_{\e_8}\tau_{\e_0-\e_i-\e_8},& \nonumber\\
& \mbox{(C)}^{\rm i}_0 &&
[\ep_{jk}]\tau_{\e_8}\tau_{2\e_0-\e_i-\e_j-\e_k-\e_8-\e_9}
=[\ep_{ik8}]\tau_{\e_0-\e_k-\e_9}\tau_{\e_0-\e_i-\e_j}
-[\ep_{ij8}]\tau_{\e_0-\e_j-\e_9}\tau_{\e_0-\e_i-\e_k},&\nonumber\\
& \mbox{(C)}^{\rm r}_0 &&
[\ep_{ij}]\tau_{\e_0-\e_i-\e_j}\tau_{\e_0-\e_k-\e_9}
+(i,j,k)\mbox{-cyclic}=0,&\nonumber\\
& \mbox{(D)}_0 &&
[\ep_{jk}]\tau_{\e_7}\tau_{\e_0-\e_i-\e_7}
=[\ep_{ik7}]\tau_{\e_j}\tau_{\e_0-\e_i-\e_j}
-[\ep_{ij7}]\tau_{\e_k}\tau_{\e_0-\e_i-\e_k} &
 \label{bi_n=0}
\end{alignat}
for mutually distinct indices $i,j,k\in C$. The bilinear equations (\ref{bi_2}) are also classif\/ied in a similar way into four types, each of which we denote by (A)${}'_n$, (B)${}'_n$, (C)${}'_n$ and (D)${}'_n$ to distinguish them from the bilinear equations (\ref{bi_1}). Typical equations are given by
\begin{alignat}{3}
& \mbox{(A)}'_0\quad &&
 [\ep_{78}][\dl-\ep_{569}]\tau_{\e_9}\tau_{2\e_0-\e_1-\e_2-\e_3-\e_4-\e_9}
+(7,8,9)\mbox{-cyclic}=0,&\nonumber\\
& \mbox{(B)}'_0 &&
[\ep_{ij}][\ep_{kl}]\tau_{\e_8}\tau_{2\e_0-\e_i-\e_j-\e_k-\e_l-\e_8}& \nonumber\\
&&& \qquad{}=[\ep_{il8}][\ep_{jk8}]\tau_{\e_0-\e_i-\e_k}\tau_{\e_0-\e_j-\e_l}
 -[\ep_{jl8}][\ep_{ik8}]\tau_{\e_0-\e_j-\e_k}\tau_{\e_0-\e_i-\e_l},& \nonumber\\
& \mbox{(C)}'_0 &&
[\ep_{ij}][\ep_{ij9}]\tau_{\e_k}\tau_{\e_0-\e_k-\e_9} +(i,j,k)\mbox{-cyclic}=0,&\nonumber\\
& \mbox{(D)}'_0 && [\ep_{ij}][\ep_{kl}]\tau_{ij}\tau_{kl}+(i,j,k)\mbox{-cyclic}=0, \qquad [\ep_{ij}][\ep_{ij7}]\tau_k\tau_{k7}+(i,j,k)\mbox{-cyclic}=0 &
 \label{bi'_n=0}
\end{alignat}
for mutually distinct indices $i,j,k,l\in C$. The bilinear equations (\ref{bi_3}) are also classif\/ied into the type (A)$^{\rm d}_n$, (B)$^{\rm d}_n$, (C)$^{\rm d}_n$ and (D)$^{\rm d}_n$. Typical equations are given by
\begin{alignat}{3}
& \mbox{(A)}^{\rm d}_0\quad &&
\tau_{\e_8}\tau_{2\e_0-\e_1-\e_2-\e_3-\e_4-\e_8}-
\tau_{\e_9}\tau_{2\e_0-\e_1-\e_2-\e_3-\e_4-\e_9}
+[\dl-\ep_{567}][\ep_{89}]\,d_C\tau_{\e_5}\tau_{\e_6}=0,& \nonumber\\
& \mbox{(B)}^{\rm d}_0 &&
\tau_{\e_i}\tau_{\e_0-\e_i-\e_7}-\tau_{\e_j}\tau_{\e_0-\e_j-\e_7}
+[\ep_{ij}][\ep_{ij7}]\,d_L\tau_{\e_8}\tau_{\e_9}=0,&\nonumber\\
&&&
\tau_{\e_8}\tau_{2\e_0-\e_1-\e_2-\e_3-\e_4-\e_8}-
\tau_{\e_0-\e_1-\e_2}\tau_{\e_0-\e_3-\e_4}
-[\ep_{128}][\ep_{348}]\,d_C\tau_{\e_5}\tau_{\e_6}=0,&\nonumber\\
& \mbox{(C)}^{\rm d}_0&&
\tau_{\e_i}\tau_{\e_0-\e_i-\e_9}-\tau_{\e_j}\tau_{\e_0-\e_j-\e_9}
+[\ep_{ij}][\ep_{ij9}]\,d_L\tau_{\e_7}\tau_{\e_8}=0,&\nonumber\\
& \mbox{(D)}^{\rm d}_0&&
\tau_{\e_0-\e_1-\e_4}\tau_{\e_0-\e_2-\e_3}-
\tau_{\e_0-\e_1-\e_3}\tau_{\e_0-\e_2-\e_4}
+[\ep_{12}][\ep_{34}]\,d_C\tau_{\e_5}\tau_{\e_6}=0&
 \label{bi''_n=0}
\end{alignat}
for mutually distinct indices $i,j\in C$.

\begin{lemma}
Any bilinear equation of type ${\rm (A)}_0$ can be obtained by an action of $\wt{W}\big(D_6^{(1)}\big)$ on the first equation of~\eqref{bi_n=0}. Also, we have a similar situation regarding each case of type ${\rm (B)}_0$, ${\rm (C)}^{\rm r}_0$, ${\rm (C)}^{\rm i}_0$ and ${\rm (D)}_0$, respectively.
\end{lemma}

\begin{proof}
Any lattice $\tau$-function on $M_1$ can be transformed to $\tau_{\e_8}$ by an action of $\wt{W}\big(D_6^{(1)}\big)$. Searching for $\Lam\in M_{-1}$ such that $\br{\Lam+\e_8,\Lam+\e_8}=0$, we f\/ind that the lattice $\tau$-functions on $M_{-1}$ which can pair with $\tau_{\e_8}$ are $\tau_{\e_0-\e_i-\e_8},\tau_{2\e_0-\e_i-\e_j-\e_k-\e_7-\e_8}$ and $\tau_{c+\e_i+\e_9-\e_7}$ for mutually distinct indices $i,j,k\in C$. Any of them can be transformed to $\tau_{\e_0-\e_i-\e_8}$ by an action of $W(D_6)$. Since $\tau_{\e_8}$ is invariant under the action of $W(D_6)$, we f\/ind that one of the pairs of the lattice $\tau$-functions in a bilinear equation of type (A)$_0$ can be transformed to $\tau_{\e_8}\tau_{\e_0-\e_i-\e_8}$ by an action of $\wt{W}\big(D_6^{(1)}\big)$. Note that three pairs of the lattice $\tau$-functions in a bilinear equation have a common barycenter. Therefore, the bilinear equations of type (A)$_0$ including the term $\tau_{\e_8}\tau_{\e_0-\e_i-\e_8}$ are reduced to
\begin{gather*}
[\ep_{89}]\tau_{\e_j}\tau_{\e_0-\e_i-\e_j}
=[\ep_{ij9}]\tau_{\e_8}\tau_{\e_0-\e_i-\e_8}
-[\ep_{ij8}]\tau_{\e_9}\tau_{\e_0-\e_i-\e_9},\\
[\ep_{89}]\tau_{\e_7}\tau_{\e_0-\e_i-\e_7}
=[\ep_{79}]\tau_{\e_8}\tau_{\e_0-\e_i-\e_8}
-[\ep_{78}]\tau_{\e_9}\tau_{\e_0-\e_i-\e_9},
\end{gather*}
which are transformed by the action of the Dynkin diagram automorphism $\rho\in\wt{W}\big(D_6^{(1)}\big)$ to each other. The proof for the other types of bilinear equations is given in a similar way.
\end{proof}

From this lemma and similar consideration for the bilinear equations (\ref{bi'_n=0}) and (\ref{bi''_n=0}), we immediately get the following proposition.

\begin{proposition}
Fix $n\in\BZ$.
\begin{enumerate}\itemsep=0pt
\item[{\rm 1.}] All the bilinear equations of type ${\rm (A)}_n$ can be transformed by the action of $\wt{W}\big(D_6^{(1)}\big)$ to one another. Also, we have a similar situation regarding each case of type ${\rm (B)}_n$, ${\rm (C)}^{\rm i}_n$, ${\rm (C)}^{\rm r}_n$ and ${\rm (D)}_n$, respectively.

\item[{\rm 2.}] All the bilinear equations of type ${\rm (A)}'_n$ can be transformed by the action of $\wt{W}\big(D_6^{(1)}\big)$ to one another. Also, we have a similar situation regarding each case of type ${\rm (B)}'_n$ and ${\rm (C)}'_n$, respectively. The set of all the bilinear equations of type ${\rm (D)}'_n$ is decomposed to two orbits by the action of $\wt{W}\big(D_6^{(1)}\big)$.

\item[{\rm 3.}] All the bilinear equations of type ${\rm (A)}^{\rm d}_n$ can be transformed by the action of $\wt{W}\big(D_6^{(1)}\big)$ to one another. Also, we have a similar situation regarding each case of type ${\rm (C)}^{\rm d}_n$ and ${\rm (D)}^{\rm d}_n$, respectively. The set of all the bilinear equations of type ${\rm (B)}^{\rm d}_n$ is decomposed to two orbits by the action of $\wt{W}\big(D_6^{(1)}\big)$.
\end{enumerate}
\end{proposition}

Let us discuss the relationships among the above types of bilinear equations.

\begin{proposition}\label{consistency_1}
If the lattice $\tau$-functions satisfy all the bilinear equations of type ${\rm (B)}_n$, then they also satisfy those of type ${\rm (A)}_n$; that is,
\[
1.\quad {\rm (B)}_n \ \Ra \ {\rm (A)}_n.
\]
Similarly, we have
\[
2.\quad {\rm (C)}^{\rm i}_n \ \Ra \ {\rm (C)}^{\rm r}_n.
\]
Moreover, if $\tau_{\Lam}\ne 0$ for $\Lam\in M_{n-1}$, we have the following:
\begin{gather*}
3.\quad {\rm (C)}^{\rm i}_{n-1} \ \Ra \ {\rm (D)}_n.\\
4.\quad {\rm (A)}_n,\ {\rm (C)}^{\rm i}_{n-1} \ \Ra \ {\rm (C)}^{\rm i}_n.
\end{gather*}
\end{proposition}

\begin{proof}
It is suf\/f\/icient to verify the statement for a certain $n\in\BZ$. The f\/irst and second statements are easily verif\/ied for the case of $n=0$.

3. ${\rm (C)}^{\rm i}_0 \ \Ra \ {\rm (D)}_1$. Let us consider the following bilinear equation
\[
[\ep_{23}]\tau_{\e_4}\tau_{2\e_0-\e_2-\e_3-\e_4-\e_5-\e_9}
=[\ep_{349}]\tau_{\e_0-\e_2-\e_9}\tau_{\e_0-\e_3-\e_5}
-[\ep_{249}]\tau_{\e_0-\e_3-\e_9}\tau_{\e_0-\e_2-\e_5}
\]
of type (C)$^{\rm i}_0$. Multiplying this equation by $\tau_{\e_0-\e_1-\e_9}$ and summing up its $(1,2,3)$-cyclic permutations, we get a bilinear equation of type (D)$_1$.

4. ${\rm (A)}_0$ and ${\rm (C)}^{\rm i}_{-1} \ \Ra \ {\rm (C)}^{\rm i}_0$. Let us consider the following bilinear equation of type (C)$^{\rm i}_{-1}$
\[
[\ep_{jk}]\tau_{\e_9}\tau_{2\e_0-\e_i-\e_j-\e_k-\e_8-\e_9}
=[\ep_{ik9}]\tau_{\e_0-\e_k-\e_8}\tau_{\e_0-\e_i-\e_j}
-[\ep_{ij9}]\tau_{\e_0-\e_j-\e_8}\tau_{\e_0-\e_i-\e_k}.
\]
Multiplying both right and left-hand sides by $\tau_{\e_8}$ and using the f\/irst equation of (\ref{bi_n=0}), we get
\begin{gather*}
\tau_{\e_9}\times
[\ep_{jk}]\tau_{\e_8}\tau_{2\e_0-\e_i-\e_j-\e_k-\e_8-\e_9}\\
\qquad {}=\tau_{\e_0-\e_i-\e_j}\times
\big( [\ep_{ik8}]\tau_{\e_9}\tau_{\e_0-\e_k-\e_9}
      +[\ep_{89}]\tau_{\e_i}\tau_{\e_0-\e_i-\e_k}\big)
-\left\{ j \lra k\right\}\\[2mm]
\qquad{} =\tau_{\e_9}\times
\big([\ep_{ik8}]\tau_{\e_0-\e_k-\e_9}\tau_{\e_0-\e_i-\e_j}
    -[\ep_{ij8}]\tau_{\e_0-\e_j-\e_9}\tau_{\e_0-\e_i-\e_k}\big),
\end{gather*}
which is equivalent to the third equation of (\ref{bi_n=0}).
\end{proof}

Also, by not dif\/f\/icult but tedious procedure, we get the following propositions.

\begin{proposition}\label{consistency_2}
If the lattice $\tau$-functions satisfy all the bilinear equations of type ${\rm (B)}^{\rm d}_n$, then they also satisfy those of type ${\rm (A)}^{\rm d}_n$; that is,
\[
1.\quad {\rm (B)}^{\rm d}_n \ \Ra \ {\rm (A)}^{\rm d}_n.
\]
Similarly, if $\tau_{\Lam}\ne 0$ for $\Lam\in M_{n-1}$, we have the following:
\begin{gather*}
2.\quad {\rm (A)}_n,\ {\rm (C)}^{\rm d}_{n-1} \ \Ra \ {\rm (C)}^{\rm d}_n,\\
3.\quad {\rm (C)}^{\rm i}_{n-1},\ {\rm (C)}^{\rm d}_{n-1} \ \Ra \ {\rm (D)}^{\rm d}_n,\\
4.\quad {\rm (C)}^{\rm d}_{n-1},\ {\rm (C)}^{\rm i}_{n-1},\ {\rm (B)}_n \ \Ra \ {\rm (B)}^{\rm d}_n.
\end{gather*}
\end{proposition}

\begin{proposition}\label{consistency_3}
If the lattice $\tau$-functions satisfy all the bilinear equations of type ${\rm (C)}^{\rm i}_n$, then they also satisfy those of type ${\rm (C)}'_n$; that is,
\[
1.\quad {\rm (C)}^{\rm i}_n  \ \Ra \ {\rm (C)}'_n.
\]
Similarly, we have
\[
2.\quad {\rm (B)}'_n \ \Ra \ {\rm (A)}'_n.
\]
Moreover, if $\tau_{\Lam}\ne 0$ for $\Lam\in M_{n-1}$, we have the following:
\begin{gather*}
3.\quad {\rm (C)}'_{n-1}, \ {\rm (D)}_n \ \Ra \ {\rm (D)}'_n,\\
4.\quad {\rm (B)}'_n, \ {\rm (C)}^{\rm i}_{n-1} \ \Ra \ {\rm (B)}_n.
\end{gather*}
\end{proposition}

\section[The construction of the $\tau$-functions on $M_0$]{The construction of the $\boldsymbol{\tau}$-functions on $\boldsymbol{M_0}$}\label{M0}

Hereafter, we construct the hypergeometric $\tau$-functions for the $q$-Painlev\'e system of type $E_7^{(1)}$ by imposing the following boundary condition
\begin{equation}
\tau_{\Lam_{-1}}=0\qquad\mbox{for any}
\quad\Lam_{-1}\in M_{-1}\label{Rc_tau}
\end{equation}
and $\tau_{\Lam_0}\ne 0$ for any $\Lam_0\in M_0$. In this section, we discuss the construction of the $\tau$-functions on the lattice $M_0$.

First, let us consider the following bilinear equations of type
(A)$_0$, (A)${}'_0$ and (A)$^{\rm d}_0$
\begin{gather}
[\ep_{89}]\tau_{\e_j}\tau_{\e_0-\e_i-\e_j}
=[\ep_{ij9}]\tau_{\e_8}\tau_{\e_0-\e_i-\e_8}
-[\ep_{ij8}]\tau_{\e_9}\tau_{\e_0-\e_i-\e_9}\qquad (i,j\in C),\nonumber\\
 [\ep_{78}][\dl-\ep_{569}]\tau_{\e_9}\tau_{2\e_0-\e_1-\e_2-\e_3-\e_4-\e_9}
+(7,8,9)\mbox{-cyclic}=0,\nonumber\\
[\dl-\ep_{567}][\ep_{89}] d_C\tau_{\e_5}\tau_{\e_6}
+\tau_{\e_8}\tau_{2\e_0-\e_1-\e_2-\e_3-\e_4-\e_8}
-\tau_{\e_9}\tau_{2\e_0-\e_1-\e_2-\e_3-\e_4-\e_9}=0.
  \label{A0}
\end{gather}
The boundary condition (\ref{Rc_tau}) leads us to
\begin{equation}
[\ep_{89}]=0\quad\Lra\quad\ep_{89}=\omega\in\BZ.\label{Rc_para}
\end{equation}
All the bilinear equations of type (A)$_0$, (A)${}'_0$ and (A)$^{\rm d}_0$ hold under the conditions (\ref{Rc_tau}) and (\ref{Rc_para}), since they can be obtained by the action of $\wt{W}\big(D_6^{(1)}\big)=\br{s_{127},s_{12},\ldots,s_{56},s_{\dl-\ep_{567}},\rho}$ on (\ref{A0}) and the coef\/f\/icient $[\ep_{89}]$ is $\wt{W}\big(D_6^{(1)}\big)$-invariant.

Under the boundary condition (\ref{Rc_tau}), the bilinear equations of type (B)$_0$, (B)${}'_0$ and (B)$^{\rm d}_0$ are expressed in terms of the lattice $\tau$-functions on $M_0$. Typical equations of these types are given~by
\begin{gather*}
[\ep_{78}]\tau_{\e_j}\tau_{\e_0-\e_i-\e_j}
=[\ep_{ij8}]\tau_{\e_7}\tau_{\e_0-\e_i-\e_7}
-[\ep_{ij7}]\tau_{\e_8}\tau_{\e_0-\e_i-\e_8},\\
[\ep_{ij}][\ep_{kl}]\tau_{\e_8}\tau_{2\e_0-\e_i-\e_j-\e_k-\e_l-\e_8}
=[\ep_{il8}][\ep_{jk8}]\tau_{\e_0-\e_i-\e_k}\tau_{\e_0-\e_j-\e_l}
-[\ep_{jl8}][\ep_{ik8}]\tau_{\e_0-\e_j-\e_k}\tau_{\e_0-\e_i-\e_l},\\
\tau_{\e_i}\tau_{\e_0-\e_i-\e_7}-\tau_{\e_j}\tau_{\e_0-\e_j-\e_7}
+[\ep_{ij}][\ep_{ij7}]\,d_L\tau_{\e_8}\tau_{\e_9}=0,\\
\tau_{\e_8}\tau_{2\e_0-\e_1-\e_2-\e_3-\e_4-\e_8}
-\tau_{\e_0-\e_1-\e_2}\tau_{\e_0-\e_3-\e_4}
=[\ep_{128}][\ep_{348}]\,d_C\tau_{\e_5}\tau_{\e_6}
\end{gather*}
for mutually distinct indices $i,j,k,l\in C$. These are reduced to
\begin{gather}
[\ep_{79}]\tau_{\e_j}\tau_{\e_0-\e_i-\e_j}
=[\ep_{ij9}]\tau_{\e_7}\tau_{\e_0-\e_i-\e_7},\qquad
\tau_{\e_i}\tau_{\e_0-\e_i-\e_7}=\tau_{\e_j}\tau_{\e_0-\e_j-\e_7},\nonumber\\
\tau_{\e_0-\e_1-\e_2}\tau_{\e_0-\e_3-\e_4}
+[\ep_{129}][\ep_{349}]\,d_C\tau_{\e_5}\tau_{\e_6}=0
  \label{B0}
\end{gather}
and
\begin{equation}
 [\ep_{il9}][\ep_{jk9}]\tau_{\e_0-\e_i-\e_k}\tau_{\e_0-\e_j-\e_l}
=[\ep_{jl9}][\ep_{ik9}]\tau_{\e_0-\e_j-\e_k}\tau_{\e_0-\e_i-\e_l}\label{B0_2}
\end{equation}
due to the conditions (\ref{Rc_tau}) and (\ref{Rc_para}). Obviously, the equation (\ref{B0_2}) can be derived from the third equation of (\ref{B0}) and its $\DS_6$-transforms. Also, it is not dif\/f\/icult to see that all the bilinear equations of type (D)$_0$, (D)${}'_0$ and (D)$^{\rm d}_0$ can be derived from the equations (\ref{B0}) and their $\wt{W}\big(D_6^{(1)}\big)$-transforms. Then, it is suf\/f\/icient to consider the equations (\ref{B0}) and their $\wt{W}\big(D_6^{(1)}\big)$-transforms for constructing the hypergeometric $\tau$-functions on $M_0$.

Let us consider a pair of non-zero meromorphic functions $\left(G(x),F(x)\right)$ satisfying the dif\/fe\-ren\-ce equations $G(x+\dl)=\epsilon\,[x]\,G(x)$ and $F(x+\dl)=G(x)F(x)$ with a constant $\epsilon\in\BC^*$. When $\mbox{Im}\,\dl>0$, a typical choice of such functions is given by
\[
G(x)=\frac{e\left(-\frac{\dl}{2}\binom{x/\dl}{2}\right)}{(u;q)_{\infty}},\qquad
F(x)=e\left(-\tfrac{\dl}{2}{\textstyle \binom{x/\dl}{3}}\right) (u;q,q)_{\infty},
\]
where $u=e(x)$, $q=e(\dl)$, $(u;q,q)=\prod\limits_{i,j=0}^{\infty}(1-uq^{i+j})$ and $\epsilon=-1$. For other choices of $\left(G(x),F(x),\epsilon\right)$, see Appendix in~\cite{Masuda}. In what follows, we f\/ix a triplet $\left(G_+(x),F_+(x),\epsilon_+\right)$ with a constant factor $\epsilon_+$, namely we have
\begin{equation}
G_+(x+\dl)=\epsilon_+[x]\,G_+(x),\qquad
F_+(x+\dl)=G_+(x)F_+(x).\label{eq_FG+}
\end{equation}
Also, we introduce a pair of functions $\left(G_-(x),F_-(x)\right)$ by the relations
\begin{equation}
F_-(x)=F_+(2\dl+\omega-x),\qquad G_-(x)G_+(\dl+\omega-x)=1.\label{rel_FG_pm}
\end{equation}
Note that these functions satisfy the dif\/ference equations
\begin{equation}
G_-(x+\dl)=\epsilon_-[x]\,G_-(x),\qquad F_-(x+\dl)=G_-(x)F_-(x)\label{eq_FG-}
\end{equation}
with $\epsilon_-=(-1)^{\omega+1}\epsilon_+$.

Moreover, we consider a triplet of functions $(\CA_+(x),\CB_+(x),\CC_+(x))$ def\/ined by the dif\/ference equations
\begin{gather}
\dfrac{\CA_+(x+\dl)\CA_+(x-\dl)}{\CA_+(x)\CA_+(x)}
=e(\aa x+\Da),\nonumber\\
\dfrac{\CB_+(x+\dl)\CB_+(x-\dl)}{\CB_+(x)\CB_+(x)}
=e(\aa x+\Db),\nonumber\\
\dfrac{\CC_+(x+\dl)\CC_+(x-\dl)}{\CC_+(x)\CC_+(x)}
=e(-\aa x+\Dc),
  \label{eq_abc}
\end{gather}
where $\Da$, $\Db$, $\Dc$ and $\aa$ are the complex constants satisfying $e(2\aa\dl+4\Db+2\Dc)=(-1)^{\omega+1}$ and $(-1)^{\omega+1}\epsilon_+^2e(\aa\omega+2\Da)+d_C\,e(5\Db+3\Dc)=0$. A typical example of such functions is given by $\CA_+(x)=e(\dl\aa\binom{x/\dl+1}{3}+\Da\binom{x/\dl}{2})$. Also, we introduce the functions $\CA_-(x)$, $\CB_-(x)$ and $\CC_-(x)$ by the relations
\begin{gather}
\CA_-(x)=\CA_+(2\dl+\omega-x),\qquad
\CB_-(x)=\CB_+(2\dl-x),\qquad
\CC_-(x)=\CC_+(2\dl-x).\label{rel_pm}
\end{gather}

\begin{definition}
For each $\Lam_0\in M_0$, we def\/ine the twelve functions $\tau^{(a;\pm)}_{\Lam_0}(\ep)\,(a\in C)$ by
\begin{gather}
\tau_{\Lam_0}^{(a;\pm)}(\ep) =\disp
F_{\pm}(\ep_{79}+(\br{\e_{79},\Lam_0}+1)\dl)
\prod_{i,j\in C;\,i<j}
F_{\pm\ka^{(a)}_{ij}}(\ep_{ij9}+(\br{\e_{ij9},\Lam_0}+1)\dl)\nonumber\\
\phantom{\tau_{\Lam_0}^{(a;\pm)}(\ep) =}{} \times\disp
\CA_{\pm}(\ep_{79}+(\br{\e_{79},\Lam_0}+1)\dl)
\prod_{i,j\in C;\,i<j}
\CA_{\pm\ka^{(a)}_{ij}}(\ep_{ij9}+(\br{\e_{ij9},\Lam_0}+1)\dl)\nonumber\\
\phantom{\tau_{\Lam_0}^{(a;\pm)}(\ep) =}{} \times\disp
\prod_{i\in C_a}
\CB_{\pm}(\ep_{ia7}+(\br{\e_{ia7},\Lam_0}+1)\dl)\,
\CB_{\pm}(\ep_{ia}+(\br{\e_{ia},\Lam_0}+1)\dl)\nonumber\\
\phantom{\tau_{\Lam_0}^{(a;\pm)}(\ep) =}{} \times\CC_{\pm}(\ep_{aa7}+(\br{\e_{aa7},\Lam_0}+1)\dl),
  \label{tau_0}
\end{gather}
where $\ka^{(a)}_{ij}$ is the sign factor def\/ined by $\ka^{(a)}_{ij}=(-1)^{\sharp(\{i,j\}\cap\{a\})}$ and $C_a=C\bs\{a\}$.
\end{definition}

\begin{theorem}\label{act_on_tau0}
The action of $\wt{W}\big(D_6^{(1)}\big)$ on the functions $\tau^{(a;\pm)}_{\Lam_0}(\ep)$ is described as follows:
\begin{enumerate}\itemsep=0pt
\item[\rm 1.] For any translation operator $T\in\wt{W}\big(D_6^{(1)}\big)$, we have $\tau^{(a;\pm)}_{T.\Lam_0}(\ep)=\tau^{(a;\pm)}_{\Lam_0}(T(\ep))$.

\item[\rm 2.] For any permutation $\sig\in\DS_6$, we have $\tau^{(\sig(a);\pm)}_{\sig.\Lam_0}(\ep)=\tau^{(a;\pm)}_{\Lam_0}(\sig(\ep))$.

\item[\rm 3.] Take two mutually distinct indices $i,j\in C$.
\begin{enumerate}\itemsep=0pt
\item[$(a)$] If $a\notin\{i,j\}$, then $\tau^{(a;\pm)}_{s_{ij7}.\Lam_0}(\ep)=\tau^{(a;\pm)}_{\Lam_0}(s_{ij7}(\ep))$.

\item[$(b)$] If $a\in\{i,j\}$, then $\tau^{(a;\pm)}_{s_{ij7}.\Lam_0}(\ep)=\tau^{(b;\mp)}_{\Lam_0}(s_{ij7}(\ep))$, where $b$ is an index such that $\{a,b\}=\{i,j\}$.
\end{enumerate}

\item[\rm 4.] The action of the central element $w_c\in W(D_6)$ is given by $\tau^{(a;\mp)}_{w_c.\Lam_0}(\ep)=\tau^{(a;\pm)}_{\Lam_0}(w_c(\ep))$.
\end{enumerate}
\end{theorem}

\begin{proof}
The f\/irst and second statements are obvious from the def\/inition of $\tau^{(a;\pm)}_{\Lam_0}(\ep)$. The third statement is guaranteed by the relations (\ref{rel_FG_pm}) and (\ref{rel_pm}). Since we have
\[
w_c :\ \ep_{79}\mapsto\dl+\omega-\ep_{79},\qquad
\ep_{ij9}\mapsto\dl+\omega-\ep_{ij9} \quad (i,j\in C),
\]
one can verify the fourth statement by using the relations (\ref{rel_FG_pm}) and (\ref{rel_pm}).
\end{proof}

\begin{remark}
The central element $w_c\in W(D_6)$ can be expressed by $w_c=s_{12}s_{127}s_{34}s_{347}s_{56}s_{567}$. It is easy to see that we have $Tw_c=w_cT^{-1}$ for any translation operator $T\in\wt{W}(D_6^{(1)})$.
\end{remark}

Let $S$ be a label set def\/ined by $S=\{(a\,;\,\epsilon)\,|\,a\in C,\,\epsilon=\pm 1\}$. By using the dif\/ference equa\-tions (\ref{eq_FG+}),~(\ref{eq_FG-}) and (\ref{eq_abc}), one can verify that the family of functions $\{\tau^{(\eta)}_{\Lam_0}(\ep)\}_{\Lam_0\in M_0}$ for each label $\eta\in S$ satisf\/ies the bilinear equations (\ref{B0}). Also, the set of the functions $\{\tau^{(\eta)}_{\Lam_0}(\ep)\,|\,\eta\in S$, $\Lam_0\in M_0\}$ is consistent with respect to the action of $\wt{W}\big(D_6^{(1)}\big)$ in the sense of Theorem~\ref{act_on_tau0}. Then, we have the following theorem.

\begin{theorem}\label{tau0_bi}
For each label $\eta\in S$, the family of functions $\{\tau_{\Lam_0}^{(\eta)}(\ep)\}_{\Lam_0\in M_0}$ defined by \eqref{tau_0} satisfies all the bilinear equations of type ${\rm (B)}_0$, ${\rm (B)}'_0$, ${\rm (B)}^{\rm d}_0$, ${\rm (D)}_0$, ${\rm (D)}'_0$ and ${\rm (D)}^{\rm d}_0$ under the conditions~\eqref{Rc_tau} and~\eqref{Rc_para}.
\end{theorem}

Before discussing the construction of the hypergeometric $\tau$-functions on $M_n$ for $n\in\BZ_{\ge 1}$, we mention those on $M_n$ for $n\in\BZ_{<0}$.

\begin{lemma}
For any fixed $n\in\BZ_{<0}$, we have $\tau_{\Lam_n}(\ep)=0$ for any $\Lam_n\in M_n$ under the conditions~\eqref{Rc_tau} and~\eqref{Rc_para}.
\end{lemma}

\section[The construction of the $\tau$-functions on $M_1$]{The construction of the $\boldsymbol{\tau}$-functions on $\boldsymbol{M_1}$}\label{M1}

In this section, we construct the hypergeometric $\tau$-functions on $M_1$. We f\/ind that a class of bilinear equations for the lattice $\tau$-functions yields the contiguity relations for the $q$-hypergeometric function ${}_8W_7$~\cite{IR,GR_2nd}. As is well-known, the $q$-hypergeometric function ${}_8W_7$ possesses the $W(D_5)$-symmetry~\cite{LJ1}. From that, we can construct a set of twelve solutions corresponding to the coset $W(D_6)/W(D_5)$, and describe the action of $\wt{W}(D_6^{(1)})$ on the set of solutions.

\subsection[The $q$-hypergeometric function ${}_8W_7$ and its transformation formula]{The $\boldsymbol{q}$-hypergeometric function $\boldsymbol{{}_8W_7}$ and its transformation formula}

Fix a complex number $q$ with $0<|q|<1$. Let us consider the basic hypergeometric function ${}_8W_7={}_8W_7(a_0;a_1,\ldots,a_5;q,z)$ def\/ined by (\ref{8W7}). It is well-known that this function admits the transformation formula~\cite{IR,GR_2nd}
\begin{gather*}
{}_8W_7(a_0;a_1,a_2,a_3,a_4,a_5;q,z)
=\dfrac{\left(qa_0,\frac{qa_0}{a_4a_5},
              \frac{q^2a_0^2}{a_1a_2a_3a_4},
              \frac{q^2a_0^2}{a_1a_2a_3a_5};q\right)_{\infty}}
       {\left(\frac{qa_0}{a_4},\frac{qa_0}{a_5},
              \frac{q^2a_0^2}{a_1a_2a_3},
              \frac{q^2a_0^2}{a_1a_2a_3a_4a_5};q\right)_{\infty}}\\
\phantom{{}_8W_7(a_0;a_1,a_2,a_3,a_4,a_5;q,z)=}{} \times
{}_8W_7\left(\tfrac{qa_0^2}{a_1a_2a_3};
             \tfrac{qa_0}{a_2a_3},
             \tfrac{qa_0}{a_1a_3},
             \tfrac{qa_0}{a_1a_2},
             a_4,a_5;q,\tfrac{qa_0}{a_4a_5}\right),
\end{gather*}
which can be expressed by the following identity
\begin{gather*}
\frac{(q^2a_0^2/a_1a_2a_3a_4a_5;q)_{\infty}
        \prod\limits_{k=1}^5(qa_0/a_k;q)_{\infty}}
      {(qa_0;q)_{\infty}}\,
{}_8W_7(a_0;a_1,\ldots,a_5;q,z)\\
\qquad{}=\frac{(q^2\wa_0^2/\wa_1\wa_2\wa_3\wa_4\wa_5;q)_{\infty}
         \prod\limits_{k=1}^5(q\wa_0/\wa_k;q)_{\infty}}
       {(q\wa_0;q)_{\infty}}\,
{}_8W_7(\wa_0;\wa_1,\ldots,\wa_5;q,\wt{z})
\end{gather*}
with respect to the coordinate transformation
\begin{gather*}
\wa_0=qa_0^2/a_1a_2a_3,\\
\wa_1=qa_0/a_2a_3,\qquad\wa_2=qa_0/a_1a_3,\qquad\wa_3=qa_0/a_1a_2,\\
\wa_4=a_4,\quad\wa_5=a_5.
\end{gather*}
In this form, the function is manifestly invariant under the permutation of the parameters $a_1,\ldots,a_5$.

Assume that $\mbox{Im}\, \dl>0$. We relate the variables $a_i$ to $\ep_j$ by
\begin{equation}
a_0=e(\dl-\ep_{669}),\qquad a_i=e(\dl-\ep_{i69})\quad (i=1,2,\ldots,5),\qquad
q=e(\dl). \label{ep<->a}
\end{equation}
Since the action of $s_{457}\in W(D_6)=\br{s_{12},s_{23},s_{34},s_{45},s_{56},s_{127}}$ on the variables $a_i$ is given by
\begin{gather*}
s_{457} :~~a_0\mapsto qa_0^2/a_1a_2a_3,\\
\phantom{s_{457} :{}}~~a_1\mapsto qa_0/a_2a_3,\qquad
  a_2\mapsto qa_0/a_1a_3,\qquad
  a_3\mapsto qa_0/a_1a_2,\\
\phantom{s_{457} :{}}~~a_4\mapsto a_4,\quad a_5\mapsto a_5,
\end{gather*}
we see that this action leads us to the above transformation formula for ${}_8W_7$.

Let us introduce the function $\mu^{(6)}(\ep)$ that is invariant under the action of the symmetric group $\DS_5=\br{s_{12},s_{23},s_{34},s_{45}}\subset\DS_6$ and satisf\/ies
\begin{equation}
\dfrac{\mu^{(6)}(s_{457}(\ep))}{\mu^{(6)}(\ep)}=
\dfrac{g_+(\ep_{459})g_+(2\dl-\ep_{669})
       \prod\limits_{i=4,5}g_+(\dl+\omega-\ep_{i67})}
      {g_+(\ep_{79})g_+(2\dl-\ep_{669}-\ep_{457})
       \prod\limits_{i=4,5}g_+(\dl+\omega-\ep_{i6})},
\label{trf_mu}
\end{equation}
where $g_+(x)$ is given by $G_+(x)=\dfrac{g_+(x)}{(u;q)_{\infty}}$ with $u=e(x)$ and $q=e(\dl)$. The relation (\ref{trf_mu}) means that the function $\dfrac{g_+(2\dl-\ep_{669})}{g_+(\ep_{79})} \prod\limits_{i\in C_6}\dfrac{1}{g_+(\dl+\omega-\ep_{i6})} \mu^{(6)}(\ep)$ is invariant under the action of $s_{457}$. Then, we see that the function
\begin{equation}
\mu^{(6)}(\ep)
\frac{G_+(2\dl-\ep_{669})}{G_+(\ep_{79})}
\prod_{i\in C_6}G_-(\ep_{i6}) \Phi^{(6)}(\ep),
\label{inv_fnct}
\end{equation}
where $\Phi^{(6)}(\ep)={}_8W_7(a_0;a_1,a_2,a_3,a_4,a_5;q,z)$, is invariant under the action of the f\/inite Weyl group $W(D_5)=\br{s_{12},s_{23},s_{34},s_{45},s_{127}}\subset W(D_6)$.

\subsection[The contiguity relations for ${}_8W_7$]{The contiguity relations for $\boldsymbol{{}_8W_7}$}\label{cont}

It is also known that the $q$-hypergeometric function $\Phi^{(6)}={}_8W_7$ satisf\/ies the following contiguity relations~\cite{IR,GR_2nd}
\begin{gather}
(a_1-a_2)(1-z) \Phi^{(6)}
=a_1\frac{\prod\limits_{i=3}^5(1-qa_0/a_1a_i)}{1-qa_0/a_1}
\Phi^{(6)}|_{a_1\mapsto a_1/q}\nonumber\\
\phantom{(a_1-a_2)(1-z) \Phi^{(6)}=}{}
-a_2\dfrac{\prod\limits_{i=3}^5(1-qa_0/a_2a_i)}{1-qa_0/a_2}
\Phi^{(6)}|_{a_2\mapsto a_2/q},
   \label{cont_8W7_1a}\\
(a_2-a_1)(1-a_0/a_1a_2)\Phi^{(6)}
=(1-a_1)(1-a_0/a_1)\Phi^{(6)}|_{a_1\mapsto qa_1}\nonumber\\
\phantom{(a_2-a_1)(1-a_0/a_1a_2)\Phi^{(6)}=}{} -(1-a_2)(1-a_0/a_2) \Phi^{(6)}|_{a_2\mapsto qa_2},
 \label{cont_8W7_2a}\\
(1-a_0/a_1)(1-z)\Phi^{(6)}
=\frac{\prod\limits_{i=1}^5(1-a_0/a_i)}
       {(1-q^{-1}a_0)(1-a_0)(1-q^{-1}a_1)} \Phi^{(6)}(-)\nonumber\\
 \phantom{(1-a_0/a_1)(1-z)\Phi^{(6)}=}{} -q^{-1}a_1
\frac{\prod\limits_{i=2}^5(1-qa_0/a_1a_i)}
      {(1-q^{-1}a_1)(1-qa_0/a_1)} \Phi^{(6)}|_{ a_1\mapsto a_1/q},
  \label{cont_8W7_1b}
\\
\Phi^{(6)}|_{a_1\mapsto qa_1}-\Phi^{(6)}=
q^{-1}z
\frac{(1-qa_0)(1-q^2a_0)\prod\limits_{i=2}^5(1-a_i)}
      {(1-a_0/a_1)\prod\limits_{i=1}^5(1-qa_0/a_i)}\Phi^{(6)}(+),
\label{cont_8W7_2b}
\end{gather}
where $\Phi^{(6)}(\pm)=\Phi^{(6)}|_{a_0\mapsto q^{\pm 2}a_0,a_1\mapsto q^{\pm 1}a_1,\ldots,a_5\mapsto q^{\pm 1}a_5}$.

Noticing that the action of translation operators $\wt{T}_{i7}\in\wt{W}\big(D_6^{(1)}\big)$ $(i\in C)$ on the variables $a_i$ $(i=0,1,\ldots,5)$ is given by
\[
\wt{T}_{i7}:\ a_i\mapsto q^{-1}a_i,\qquad
\wt{T}_{67}:\ a_0\mapsto q^{-2}a_0,\qquad a_i\mapsto q^{-1}a_i
\qquad (i\in C_6),
\]
we see that the contiguity relations (\ref{cont_8W7_1a}) and (\ref{cont_8W7_2a}) can be rewritten as
\begin{gather}
(-1)^{\omega}[\ep_{jk}][\ep_{79}]\,\Phi^{(6)}(\ep)
=\frac{\prod\limits_{l\in C_6\bs\{j,k\}}[\ep_{jl9}]}
        {[\ep_{j6}-\dl]}\Phi^{(6)}\big(\wt{T}_{j7}(\ep)\big)
-\frac{\prod\limits_{l\in C_6\bs\{j,k\}}[\ep_{kl9}]}
        {[\ep_{k6}-\dl]}\Phi^{(6)}\big(\wt{T}_{k7}(\ep)\big)\label{cont_8W7_1a'}
\end{gather}
and
\begin{equation}
[\ep_{jk}][\ep_{jk9}-\dl]\,\Phi^{(6)}(\ep)
=[\ep_{j69}-\dl][\ep_{j6}]\Phi^{(6)}\big(\wt{T}_{j7}^{-1}(\ep)\big)
-[\ep_{k69}-\dl][\ep_{k6}]\Phi^{(6)}\big(\wt{T}_{k7}^{-1}(\ep)\big),\label{cont_8W7_2a'}
\end{equation}
respectively, for $j,k\in C_6$. Similarly, the contiguity relations (\ref{cont_8W7_1b}) and (\ref{cont_8W7_2b}) are expressed by
\begin{gather}
(-1)^{\omega}[\ep_{k6}][\ep_{k69}][\ep_{79}]\Phi^{(6)}(\ep)
=\frac{\prod\limits_{l\in C_6\bs\{k\}}[\ep_{kl9}]}
       {[\ep_{k6}-\dl]}\Phi^{(6)}\big(\wt{T}_{k7}(\ep)\big)\nonumber\\
\phantom{(-1)^{\omega}[\ep_{k6}][\ep_{k69}][\ep_{79}]\Phi^{(6)}(\ep)=}{}
-\frac{\prod\limits_{l\in C_6}[\ep_{l6}]}
       {[\dl-\ep_{669}][-\ep_{669}]}\Phi^{(6)}\big(\wt{T}_{67}(\ep)\big)
\label{cont_8W7_1b'}
\end{gather}
and{\samepage
\begin{equation}
\Phi^{(6)}(\ep)=\Phi^{(6)}(\wt{T}_{k7}^{-1}(\ep))
-\frac{[3\dl-\ep_{669}][2\dl-\ep_{669}]
        \prod\limits_{l\in C_6\bs\{k\}}[\ep_{l69}-\dl]}
       {[\ep_{k6}]\prod\limits_{l\in C_6}[\ep_{l6}-\dl]}
\Phi^{(6)}\big(\wt{T}_{67}^{-1}(\ep)\big),
\label{cont_8W7_2b'}
\end{equation}
respectively, for $k\in C_6$.}

Let us introduce the function $\Psi^{(6)}(\ep)$ by
\[
\prod_{i,j\in C_6;\,i<j}\dfrac{1}{G_+(\ep_{ij9})}\,
\Psi^{(6)}(\ep)=
\mu^{(6)}(\ep)\,
\dfrac{G_+(2\dl-\ep_{669})}{G_+(\ep_{79})}\,
\prod_{i\in C_6}G_-(\ep_{i6})\,\Phi^{(6)}(\ep),
\]
where the right-hand side is the $W(D_5)$-invariant function (\ref{inv_fnct}). We see that the function $\Psi^{(6)}(\ep)$ is $\DS_5$-invariant and satisf\/ies the relation
\[
\prod_{i,j\in\{1,2,3\};\,i<j}G_+(\ep_{ij9})
\prod_{i=1,2,3}G_-(\ep_{i69})\,
G_+(\ep_{459})\,\Psi^{(6)}(s_{457}(\ep))
=G_+(\ep_{79})\Psi^{(6)}(\ep).
\]
Suppose that the correction factor $\mu^{(6)}(\ep)$, introduced in the previous subsection, satisf\/ies the dif\/ference equation $\mu^{(6)}(\wt{T}_{i7}(\ep))=(-1)^{\omega}\mu^{(6)}(\ep)\,(i\in C)$. Then both of the contiguity rela\-tions~(\ref{cont_8W7_1a'}) and (\ref{cont_8W7_1b'}) yield
\[
(-1)^{\omega+1}\epsilon_+^2
[\ep_{jk}][\ep_{jk9}]\Psi^{(6)}(\ep)
=\Psi^{(6)}\big(\wt{T}_{j7}(\ep)\big)-\Psi^{(6)}\big(\wt{T}_{k7}(\ep)\big)
\]
for $j,k\in C$. Similarly, we see that (\ref{cont_8W7_2a'}) and (\ref{cont_8W7_2b'}) are reduced to
\begin{gather*}
\epsilon_+^{-2}
[\ep_{jk}][\ep_{79}-\dl]\Psi^{(6)}(\ep)
=\prod_{l\in C\bs\{j,k\}}[\ep_{kl9}-\dl]
\Psi^{(6)}\big(\wt{T}_{k7}^{-1}(\ep)\big)
-\prod_{l\in C\bs\{j,k\}}[\ep_{jl9}-\dl]
\Psi^{(6)}\big(\wt{T}_{j7}^{-1}(\ep)\big)
\end{gather*}
for $j,k\in C$. It is easy to see that the function $\Psi^{(a)}(\ep)$ $(a\in C_6)$ def\/ined by $\Psi^{(a)}(\ep)=\Psi^{(6)}(s_{a6}(\ep))$ satisf\/ies the same contiguity relations as those for $\Psi^{(6)}(\ep)$.

\begin{proposition}
Each of the functions $\Psi^{(a)}(\ep)$ $(a\in C)$ satisfies the contiguity relations
\begin{gather}
(-1)^{\omega+1}\epsilon_+^2
[\ep_{jk}][\ep_{jk9}]\Psi^{(a)}(\ep)
=\Psi^{(a)}\big(\wt{T}_{j7}(\ep)\big)
-\Psi^{(a)}\big(\wt{T}_{k7}(\ep)\big),
\label{cont_1}
\\
\epsilon_+^{-2}
[\ep_{jk}][\ep_{79}-\dl]\Psi^{(a)}(\ep)\nonumber\\
\qquad {}=\prod_{l\in C\bs\{j,k\}}[\ep_{kl9}-\dl]
\Psi^{(a)}\big(\wt{T}_{k7}^{-1}(\ep)\big)
-\prod_{l\in C\bs\{j,k\}}[\ep_{jl9}-\dl]
\Psi^{(a)}\big(\wt{T}_{j7}^{-1}(\ep)\big)
 \label{cont_2}
\end{gather}
for mutually distinct indices $j,k\in C$.
\end{proposition}

Here, we give a remark on choice of the correction factor $\mu^{(6)}(\ep)$. The function $\mu^{(6)}(\ep)$ in the form
\[
\mu^{(6)}(\ep)=\nu^{(6)}(\ep)
\frac{g_+(\ep_{79})\prod\limits_{i\in C_6}g_+(\dl+\omega-\ep_{i6})}
      {g_+(2\dl-\ep_{669})},
\]
where $\nu^{(6)}(\ep)$ is a $W(D_5)$-invariant function, is manifestly $\DS_5$-invariant and satisf\/ies the relation~(\ref{trf_mu}). Due to $g_+(x+\dl)=-\epsilon_+e(-\hf x) g_+(x)$, what we have to do is to f\/ind a $W(D_5)$-invariant function $\nu^{(6)}(\ep)$ satisfying the dif\/ference equations
\begin{gather}
\nu^{(6)}(\wt{T}_{i7}(\ep))=
(-1)^{\omega}\epsilon_+^{-2} e\left(\tfrac 12 (\ep_{79}-\ep_{i6}+\dl+\omega)\right)
\nu^{(6)}(\ep)\qquad (i\in C_6),\nonumber\\
\nu^{(6)}(\wt{T}_{67}(\ep))=
(-1)^{\omega}\epsilon_+^2 e\left(\tfrac 12 \ep_{669}\right) \nu^{(6)}(\ep).
  \label{eq_nu}
\end{gather}
It is easy to see that the function $\nu^{(6)}(\ep)$ in the form
\[
\nu^{(6)}(\ep)=
\varphi_1(\ep_{79})
\prod_{i,j\in C_6\,;\,i<j}\varphi_1(\ep_{ij9})
\prod_{i\in C_6}\varphi_1(\dl+\omega-\ep_{i69})
\prod_{i\in C_6}\varphi_2(\ep_{i67})\varphi_2(\ep_{i6})
\varphi_3(\ep_{667}),
\]
where $\varphi_i(x)$ $(i=1,2,3)$ are arbitrary functions, is $W(D_5)$-invariant. When $\varphi_i(x)$ $(i=1,2,3)$ satisfy $\varphi_i(x+\dl)=e(\aa_i x+\beta_i) \varphi_i(x)$ with $\aa_3=2\aa_1-\aa_2$, $8\aa_1+4\aa_2=1$ and $\epsilon_+^{-2}e((\aa_1-\aa_2)\dl+\aa_2\omega+(-4\beta_1+5\beta_2+\beta_3))=1$, the function $\nu^{(6)}(\ep)$ satisf\/ies the dif\/ference equations (\ref{eq_nu}). A~typical choice of them is given by $\varphi_i(x)=e(\aa_i\dl\binom{x/\dl}{2}+\beta_ix/\dl)$. It is possible to determi\-ne~$\mu^{(6)}(\ep)$ according to the choice of the functions $\varphi_i(x)$ $(i=1,2,3)$ and $G_+(x)$. We have proposed some examples of the functions $G_+(x)$ and $F_+(x)$ in Appendix of \cite{Masuda}.

\subsection{Twelve solutions} \label{12}

Hereafter, we denote $\Psi^{(a)}(\ep)$ by $\Psi^{(a;+)}(\ep)$. Since the action of the central element $w_c\in W(D_6)$ on the variables $a_i$ $(i=0,1,\ldots,5)$ is given by $w_c(a_i)=q/a_i$, the application of $w_c$ to the contiguity relations (\ref{cont_1}) and (\ref{cont_2}) leads us to
\begin{gather*}
\epsilon_+^2[\ep_{jk}][\ep_{jk9}-\dl]
\ch{\Psi}^{(a;+)}(\ep)
=\ch{\Psi}^{(a;+)}\big(\wt{T}_{k7}^{-1}(\ep)\big)
-\ch{\Psi}^{(a;+)}\big(\wt{T}_{j7}^{-1}(\ep)\big),\\
\disp
(-1)^{\omega}\epsilon_+^{-2}[\ep_{jk}][\ep_{79}]
\ch{\Psi}^{(a;+)}(\ep)
=\prod_{l\in C\bs\{j,k\}}[\ep_{kl9}]
\ch{\Psi}^{(a;+)}\big(\wt{T}_{k7}(\ep)\big)
-\prod_{l\in C\bs\{j,k\}}[\ep_{jl9}]
\ch{\Psi}^{(a;+)}\big(\wt{T}_{j7}(\ep)\big),
\end{gather*}
where $\ch{\Psi}^{(a;+)}(\ep)=\Psi^{(a;+)}(w_c(\ep))$. Let us introduce the function $\Psi^{(a;-)}(\ep)$ by
\begin{gather*}
\Psi^{(a;-)}(\ep)=
\CE^{(a;+)}(\ep) \CG^{(a;+)}(\ep) \ch{\Psi}^{(a;+)}(\ep),\\
\CE^{(a;+)}(\ep)=
\CA'_+(\ep_{79})
\prod_{i,j\in C_a;\,i<j}\CA'_+(\ep_{ij9})
\prod_{i\in C_a}\CA'_+(\dl+\omega-\ep_{ia9})
\prod_{i\in C_a}\CB'_+(\ep_{ia7})\CB'_+(\ep_{ia})
\CC'_+(\ep_{aa7}),\\
\CG^{(a;+)}(\ep)=
\dfrac{\prod\limits_{i\in C_a}G_+(\ep_{ia9})
       \prod\limits_{i,j\in C_a;\,i<j}G_-(\ep_{ij9})}
      {G_-(\ep_{79})},
\end{gather*}
where the functions $\CA'_+(x)$, $\CB'_+(x)$ and $\CC'_+(x)$ are expressed in terms of $\CA_+(x)$, $\CB_+(x)$ and $\CC_+(x)$, introduced in the previous section, by
\[
\CA'_+(x)=\dfrac{\CA_+(2\dl+\omega-x)}{\CA_+(\dl+\omega-x)},\qquad
\CB'_+(x)=\dfrac{\CB_+(-x)}{\CB_+(2\dl-x)},\qquad
\CC'_+(x)=\dfrac{\CC_+(-\dl-x)}{\CC_+(3\dl-x)}.
\]
When we set $d_C=d_L=(-1)^{\omega}\epsilon_+^2$, the factors $\CE^{(a;+)}(\ep)$ and $\CG^{(a;+)}(\ep)$ satisfy the dif\/ference equations
\begin{gather*}
\CE^{(a;+)}\big(\wt{T}_{i7}(\ep)\big)
=(-1)^{\omega+1} \CE^{(a:+)}(\ep),\qquad
\CG^{(a:+)}\big(\wt{T}_{i7}(\ep)\big)
=(-1)^{\omega+1}\epsilon_+^4
\dfrac{\prod\limits_{l\in C_i}[\ep_{il9}]}
      {[\ep_{79}]}
\CG^{(a:+)}(\ep)
\end{gather*}
for $i\in C$, and we get $e(2\aa\dl+4\Db+2\Dc)=(-1)^{\omega+1}$ and $e(\aa\omega+2\Da)=e(5\Db+3\Dc)$. Thus, we f\/ind that each of the functions $\Psi^{(a;-)}(\ep)$ satisf\/ies the contiguity relations
\begin{gather*}
\epsilon_+^{-2}[\ep_{jk}][\ep_{79}-\dl]\Psi^{(a;-)}(\ep)
=\prod_{l\in C\bs\{j,k\}}[\ep_{kl9}-\dl]
\Psi^{(a;-)}\big(\wt{T}_{k7}^{-1}(\ep)\big)\\
\phantom{\epsilon_+^{-2}[\ep_{jk}][\ep_{79}-\dl]\Psi^{(a;-)}(\ep)=}{}
-\prod_{l\in C\bs\{j,k\}}[\ep_{jl9}-\dl]
\Psi^{(a;-)}\big(\wt{T}_{j7}^{-1}(\ep)\big),\\
(-1)^{\omega+1}\epsilon_+^2[\ep_{jk}][\ep_{jk9}]\Psi^{(a;-)}(\ep)
=\Psi^{(a;-)}\big(\wt{T}_{j7}(\ep)\big)-\Psi^{(a;-)}\big(\wt{T}_{k7}(\ep)\big),
\end{gather*}
which are the same as those for $\Psi^{(a;+)}(\ep)$.

\begin{theorem}\label{eqs_Psi}
Each of the twelve functions $\Psi(\ep)=\Psi^{(a;\pm)}(\ep)$ gives rise to the solution of the contiguity relations
\begin{gather*}
(-1)^{\omega+1}\epsilon_+^2[\ep_{jk}][\ep_{jk9}]\Psi(\ep)
=\Psi\big(\wt{T}_{j7}(\ep)\big)-\Psi\big(\wt{T}_{k7}(\ep)\big),\\
\epsilon_+^{-2}[\ep_{jk}][\ep_{79}-\dl]\Psi(\ep)
=\prod_{l\in C\bs\{j,k\}}[\ep_{kl9}-\dl]
\Psi\big(\wt{T}_{k7}^{-1}(\ep)\big)
-\prod_{l\in C\bs\{j,k\}}[\ep_{jl9}-\dl]
\Psi\big(\wt{T}_{j7}^{-1}(\ep)\big)
\end{gather*}
for mutually distinct indices $j,k\in C$.
\end{theorem}

From these contiguity relations, one can get the $q$-hypergeometric equation of the second order. The functions $\Psi^{(a;\pm)}(\ep)$ coincide with the twelve pairwise linearly independent solutions to the $q$-hypergeometric equation constructed by Gupta and Masson~\cite{GM}.

Furthermore, we introduce the function $\CE^{(a;-)}(\ep)$ by
\begin{gather*}
\CE^{(a;-)}(\ep)=
\CA'_-(\ep_{79})
\prod_{i,j\in C_a ;\,i<j}\CA'_-(\ep_{ij9})
\prod_{i\in C_a}\CA'_-(\dl+\omega-\ep_{ia9})
\prod_{i\in C_a}\CB'_-(\ep_{ia7})\CB'_-(\ep_{ia})
\CC'_-(\ep_{aa7}),
\end{gather*}
where $\CA'_-(x)$, $\CB'_-(x)$ and $\CC'_-(x)$ are def\/ined by $\CA'_-(x)\CA'_+(\dl+\omega-x)=1$, $\CB'_-(x)\CB'_+(-x)=1$ and $\CC'_-(x)\CC'_+(-x)=1$, respectively. By construction, we have the following proposition.

\begin{proposition}\label{act_on_Psi}
The action of $W(D_6)$ on the functions $\Psi^{(a;\pm)}(\ep)$ is described  as follows:
\begin{enumerate}\itemsep=0pt
\item[{\rm 1.}] For any permutation $\sig\in\DS_6$, we have $\Psi^{(a;\pm)}(\sig(\ep))=\Psi^{(\sig(a);\pm)}(\ep)$.

\item[{\rm 2.}] Take two mutually distinct indices $i,j\in C$.
\begin{enumerate}\itemsep=0pt
\item[$(a)$] If $a\notin\{i,j\}$, then
\[
\Psi^{(a;\pm)}(s_{ij7}(\ep))=
\dfrac{G_{\pm}(\ep_{79})}
      {G_{\pm}(\ep_{ij9})
       \prod\limits_{\stackrel{k,l\in C\bs\{i,j,a\};}{k<l}}G_{\pm}(\ep_{kl9})
       \prod\limits_{k\in C\bs\{i,j,a\}}G_{\mp}(\ep_{ka9})}
\Psi^{(a;\pm)}(\ep).
\]

\item[$(b)$] If $a\in\{i,j\}$, then
\[
\Psi^{(a;\pm)}(s_{ij7}(\ep))=
\dfrac{1}{\CE^{(b;\pm)}(\ep)}
\dfrac{G_{\mp}(\ep_{79})}
      {G_{\pm}(\ep_{ij9})
       \prod\limits_{k,l\in C\bs\{i,j\};\,k<l}G_{\mp}(\ep_{kl9})}
\Psi^{(b;\mp)}(\ep),
\]
where $b$ is an index such that $\{a,b\}=\{i,j\}$.
\end{enumerate}

\item[{\rm 3.}] The action of the central element $w_c\in W(D_6)$ is given by
\[
\Psi^{(a;\mp)}(\ep)=
\CE^{(a;\pm)}(\ep)
\dfrac{       \prod\limits_{i\in C_a}G_{\pm}(\ep_{ia9})
       \prod\limits_{i,j\in C_a;\,i<j}G_{\mp}(\ep_{ij9})}
      {G_{\mp}(\ep_{79})}
\Psi^{(a;\pm)}(w_c(\ep)).
\]
\end{enumerate}
\end{proposition}

The set of twelve functions $\Psi^{(a;\pm)}(\ep)$ corresponds to the coset $W(D_6)/W(D_5)$, as we will see below. Note that $|W(D_6)/W(D_5)|=12$.

\subsection[The $\tau$-functions on $M_1$]{The $\boldsymbol{\tau}$-functions on $\boldsymbol{M_1}$}

Here, we construct the functions $\tau_{\Lam_1}(\ep)$ $(\Lam_1\in M_1)$ on the basis of the discussion in the previous subsections. The bilinear equations to be considered are of type (C)$_0$, (C)${}'_0$, (C)$^{\rm d}_0$, (D)$_1$, (D)${}'_1$ and (D)$^{\rm d}_1$, since the functions $\tau_{\Lam_0}(\ep)$ $(\Lam_0\in M_0)$ are already known.

It is easy to get the following lemma.

\begin{lemma}\label{C0''}
If the lattice $\tau$-functions satisfy all the bilinear equations of type ${\rm (B)}^{\rm d}_0$ and ${\rm (C)}^{\rm d}_0$ under the boundary condition \eqref{Rc_tau}, then they also satisfy those of type ${\rm (C)}^{\rm i}_0$.
\end{lemma}

From this lemma, we see that it is suf\/f\/icient for constructing the hypergeometric $\tau$-functions on $M_1$ to consider the bilinear equations of type ${\rm (C)}^{\rm d}_0$.

\begin{definition}
For each $\Lam_1\in M_1$, we def\/ine the twelve functions $\tau_{\Lam_1}^{(a;\pm)}(\ep)$ by
\begin{equation}
\tau_{\Lam_1}^{(a;\pm)}(\ep)
=\CN_{\Lam_1}^{(a;\pm)}(\ep)\,\Psi_{\Lam_1}^{(a;\pm)}(\ep),
\label{tau_1}
\end{equation}
where $\CN_{\Lam_1}^{(a;\pm)}(\ep)$ is given by
\begin{gather*}
\CN_{\Lam_1}^{(a;\pm)}(\ep)=
F_{\pm}(\ep_{79}+(\br{\e_{79},\Lam_1}+1)\dl)\disp
\prod_{i,j\in C;\,i<j}
F_{\pm\ka^{(a)}_{ij}}(\ep_{ij9}+\br{\e_{ij9},\Lam_1}\dl)\\
\phantom{\CN_{\Lam_1}^{(a;\pm)}(\ep)=}{} \times\disp
\CA_{\pm}(\ep_{79}+(\br{\e_{79},\Lam_1}+1)\dl)\\
\phantom{\CN_{\Lam_1}^{(a;\pm)}(\ep)=}{}\times\disp
\prod_{i,j\in C_a;\,i<j}
\CA_{\pm}(\ep_{ij9}+(\br{\e_{ij9},\Lam_1}+1)\dl)
\prod_{i\in C_a}
\CA_{\mp}(\ep_{ia9}+\br{\e_{ia9},\Lam_1}\dl)\\
\phantom{\CN_{\Lam_1}^{(a;\pm)}(\ep)=}{}\times\disp
\prod_{i\in C_a}
\CB_{\pm}(\ep_{ia7}+\br{\e_{ia7},\Lam_1}\dl)
\CB_{\pm}(\ep_{ia}+\br{\e_{ia},\Lam_1}\dl)\\
\phantom{\CN_{\Lam_1}^{(a;\pm)}(\ep)=}{}\times
\CC_{\pm}(\ep_{aa7}+(\br{\e_{aa7},\Lam_1}-1)\dl),
\end{gather*}
and $\Psi_{\Lam_1}^{(a;\pm)}(\ep)=\Psi^{(a;\pm)}(\ep+\br{\e,\Lam_1}\dl)$.
\end{definition}

\begin{theorem}\label{act_on_tau1}
The action of $\wt{W}\big(D_6^{(1)}\big)$ on the functions $\tau_{\Lam_1}^{(a;\pm)}(\ep)$ is described as follows:

\begin{enumerate}\itemsep=0pt
\item[{\rm 1.}] For any translation operator $T\in\wt{W}\big(D_6^{(1)}\big)$, we have $\tau^{(a;\pm)}_{T.\Lam_1}(\ep)=\tau^{(a;\pm)}_{\Lam_1}(T(\ep))$.

\item[{\rm 2.}] For any permutation $\sig\in\DS_6$, we have $\tau^{(\sig(a);\pm)}_{\sig.\Lam_1}(\ep)=\tau^{(a;\pm)}_{\Lam_1}(\sig(\ep))$.

\item[{\rm 3.}] Take two mutually distinct indices $i,j\in C$.
\begin{enumerate}
\item[$(a)$] If $a\notin\{i,j\}$, then $\tau^{(a;\pm)}_{s_{ij7}.\Lam_1}(\ep)=\tau^{(a;\pm)}_{\Lam_1}(s_{ij7}(\ep))$.

\item[$(b)$] If $a\in\{i,j\}$, then $\tau^{(a;\pm)}_{s_{ij7}.\Lam_1}(\ep)=\tau^{(b;\mp)}_{\Lam_1}(s_{ij7}(\ep))$, where $b$ is an index such that $\{a,b\}=\{i,j\}$.
\end{enumerate}

\item[{\rm 4.}] The action of the central element $w_c\in W(D_6)$ is given by $\tau^{(a;\mp)}_{w_c.\Lam_1}(\ep)=\tau^{(a;\pm)}_{\Lam_1}(w_c(\ep))$.
\end{enumerate}
\end{theorem}

\begin{proof}
The f\/irst and second statements are obvious from the def\/inition of $\tau^{(a;\pm)}_{\Lam_1}(\ep)$. The third and fourth statements are guaranteed by Proposition \ref{act_on_Psi} and (\ref{tau_1}).
\end{proof}

\begin{corollary}
For the particular element $\e_8\in M_1$, the set of twelve functions
\begin{gather*}
\tau_{\e_8}^{(a;\pm)}(\ep)=
F_{\pm}(\ep_{79}+\dl)\disp
\prod_{i,j\in C;\,i<j}F_{\pm\ka^{(a)}_{ij}}(\ep_{ij9})\\
\phantom{\tau_{\e_8}^{(a;\pm)}(\ep)=}{} \times\disp
\CA_{\pm}(\ep_{79}+\dl)
\prod_{i,j\in C_a;\,i<j}
\CA_{\pm}(\ep_{ij9}+\dl)
\prod_{i\in C_a}
\CA_{\mp}(\ep_{ia9})\\
\phantom{\tau_{\e_8}^{(a;\pm)}(\ep)=}{}\times\disp
\prod_{i\in C_a}
\CB_{\pm}(\ep_{ia7})\,\CB_{\pm}(\ep_{ia})
\times\CC_{\pm}(\ep_{aa7}-\dl) \Psi^{(a;\pm)}(\ep)
\end{gather*}
is stabilized by $W(D_6)$\footnote{Note that $\e_8\in M_1$ is $W(D_6)$-invariant.}. For each label $(a;\pm)\in S$, the isotropy subgroup of $\tau_{\e_8}^{(a;\pm)}(\ep)$ is isomorphic to $W(D_5)$;
\[
\tau_{\e_8}^{(6;\pm)}(w(\ep))=\tau_{\e_8}^{(6;\pm)}(\ep),\qquad
w\in W(D_5)=\br{s_{12},s_{23},s_{34},s_{45},s_{127}}
\]
for instance.
\end{corollary}

Let us consider the bilinear equation of type (C)$^{\rm d}_0$
\begin{equation}
\tau_{\e_i}\tau_{\e_0-\e_i-\e_9}-\tau_{\e_j}\tau_{\e_0-\e_j-\e_9}
+[\ep_{ij}][\ep_{ij9}] d_L\tau_{\e_7}\tau_{\e_8}=0\label{C0}
\end{equation}
for mutually distinct indices $i,j\in C$. Substituting (\ref{tau_0}) and (\ref{tau_1}) into (\ref{C0}), we get for $\Psi^{(a;\pm)}(\ep)$ the linear relation
\[
(-1)^{\omega+1}\epsilon_+^2[\ep_{ij}][\ep_{ij9}]\Psi^{(a;\pm)}(\ep)
=\Psi^{(a;\pm)}(\wt{T}_{i7}(\ep))
-\Psi^{(a;\pm)}(\wt{T}_{j7}(\ep)).
\]
Similarly, the application of the central element $w_c\in W(D_6)$ to the bilinear equation (\ref{C0}) leads us to
\begin{gather*}
\epsilon_+^{-2}[\ep_{jk}][\ep_{79}-\dl]\Psi^{(a;\pm)}(\ep)
=\prod_{l\in C\bs\{j,k\}}[\ep_{kl9}-\dl]
\Psi^{(a;\pm)}\big(\wt{T}_{k7}^{-1}(\ep)\big)\\
\phantom{\epsilon_+^{-2}[\ep_{jk}][\ep_{79}-\dl]\Psi^{(a;\pm)}(\ep)=}{}
-\prod_{l\in C\bs\{j,k\}}[\ep_{jl9}-\dl]
\Psi^{(a;\pm)}(\wt{T}_{j7}^{-1}(\ep))
\end{gather*}
for mutually distinct indices $i,j\in C$. These are precisely the contiguity relations in Theo\-rem~\ref{eqs_Psi}.

Also, the set of functions $\{\tau_{\Lam_1}^{(\eta)}(\ep)\,|\,\eta\in S,\Lam_1\in M_1\}$ is consistent with respect to the action of $\wt{W}\big(D_6^{(1)}\big)$ in the sense of Theorem \ref{act_on_tau1}. Therefore, we have the following theorem due to Propositions~\ref{consistency_1}, \ref{consistency_2}, \ref{consistency_3} and Lemma~\ref{C0''}.

\begin{theorem}\label{tau1_bi}
For each label $\eta\in S$, the family of functions $\{\tau_{\Lam}^{(\eta)}(x)\}_{\Lam\in M_0\coprod M_1}$ defined by~\eqref{tau_0} and \eqref{tau_1} satisfies all the bilinear equations of type ${\rm (C)}^{\rm d}_0$, ${\rm (C)}_0^{\rm i}$, ${\rm (C)}_0^{\rm r}$, ${\rm (C)}'_0$, ${\rm (D)}^{\rm d}_1$, ${\rm (D)}_1$ and ${\rm (D)}'_1$ under the conditions \eqref{Rc_tau} and \eqref{Rc_para}.
\end{theorem}

\begin{remark}
From this theorem, we see that the bilinear equations of type (C)$^{\rm d}_0$, (C)$_0^{\rm i}$, (C)$_0^{\rm r}$ and (C)${}'_0$ imply the contiguity relations for the $q$-hypergeometric function ${}_8W_7$. Also, we get the quadratic relations for ${}_8W_7$ from the bilinear equations of type (D)$^{\rm d}_1$, (D)$_1$ and (D)$'_1$.
\end{remark}

\begin{remark}
From the result in this section, one can get an explicit expression for the so-called Riccati solution to the system of $q$-dif\/ference equations (\ref{eq:E7}), in terms of the functions $\tau_{\Lam_1}^{(\eta)}(\ep)$. When the label $\eta\in S$ is f\/ixed, one can express $b_i$ $(i=1,2,\ldots,8)$ and $t$ in terms of the parameters of the $q$-hypergeometric function ${}_8W_7$. On the other hand, another expression for the Riccati solution has been proposed in~\cite{KMNOY3}, which is constructed under the condition $b_1b_3=b_5b_7$; that is, $\ep_{358}\in\BZ$. Comparing this with the condition~(\ref{Rc_para}), we f\/ind that these two expressions can be transformed to each other by a B\"acklund transformation.
\end{remark}

\section[A determinant formula for the hypergeometric $\tau$-functions]{A determinant formula for the hypergeometric $\boldsymbol{\tau}$-functions}\label{det}

One of the important features of the hypergeometric solutions to the continuous and discrete Painlev\'e equations is that they can be expressed in terms of Wronskians or Casorati determinants~\cite{HKW,KNY,KK,HK,Sakai1}. In this section, we show that the hypergeometric $\tau$-functions on $M_n$ $(n\in\BZ_{\ge 2})$ are expressed by ``two-directional Casorati determinants'' of order $n$.

Let us introduce the auxiliary variables $x_i$ $(i=0,1,\ldots,6)$ by $x_0=\dl-\ep_{78}$ and $x_i=\hf \ep_{ii9}$ $(i\in C)$, where we have $x_0+x_1+\cdots+x_6=2\dl+2\omega$. Under the conditions~(\ref{Rc_tau}) and~(\ref{Rc_para}), the functions $\tau_{\Lam}(\ep)$ depend on $x_i$ (and $\omega$). In what follows, we denote the hypergeometric $\tau$-functions by $\tau_{\Lam}(x)$ instead of by $\tau_{\Lam}(\ep)$ for convenience. Also, we denote a function $f^{(\eta)}(x)$ $(\eta\in S)$ by~$f(\eta;x)$.

For each $n\in\BZ_{\ge 0}$, we def\/ine the twelve functions $K_n(\eta;x)=K_n^{(a;\pm)}(x)$ by the following ``two-directional Casorati determinants''
\begin{gather*}
K_{2m}\left(\eta;x_0+\tfrac{2m-1}{2}\dl,x_i+\tfrac{2m-1}{4}\dl\right)
\Big|_{\,x_i\mapsto x_i-(m-1)\dl\,(i=1,2,3,4)}\\
\qquad{}=\det\big(\Psi(b-m,m+1-b,a-m,m+1-a)\big)_{a,b=1}^{2m}, \\
K_{2m+1}\left(\eta;x_0+m\dl,x_i+\tfrac{m}{2}\dl\right)
\Big|_{x_i\mapsto x_i-m\dl\,(i=1,2,3,4)}\\
\qquad{}=\det\big(\Psi(b-m-1,m+1-b,a-m-1,m+1-a)\big)_{a,b=1}^{2m+1},
\end{gather*}
where $\Psi(m_1,m_2,m_3,m_4)=\Psi(\eta;x)|_{x_i\mapsto x_i+m_i\dl \, (i=1,2,3,4)}$, and $\Psi(\eta;x)$ is the hypergeometric function multiplied by some normalization factors, introduced in Section~\ref{cont} and~\ref{12}. The f\/irst some members of $K_n(\eta;x)$ are given as follows:
\begin{gather*}
K_0(x)=1,\qquad K_1(x)=\Psi(x),\qquad
K_2\left(x_0+\tfrac 12 \dl,x_i+\tfrac 14 \dl\right)=
\left|
\begin{array}{cc}
\Psi^{24}(x)&\Psi^{14}(x)\\
\Psi^{23}(x)&\Psi^{13}(x)
\end{array}
\right|,\\
K_3\left(x_0+ \dl,x_i+\tfrac 12 \dl\right)
\Big|_{\,x_i\mapsto x_i-\dl\,(i=1,2,3,4)}=
\left|
\begin{array}{ccc}
\Psi^{24}_{13}(x)&\Psi^{4}_{3}(x)&\Psi^{14}_{23}(x)\\[1mm]
\Psi^{2}_{1}(x)&\Psi(x)&\Psi^{1}_{2}(x)\\[1mm]
\Psi^{23}_{14}(x)&\Psi^{3}_{4}(x)&\Psi^{13}_{24}(x)
\end{array}
\right|,
\end{gather*}
where we omit the label $\eta\in S$ for simplicity, and $\Psi^{i_1\ldots i_r}_{j_1\ldots j_r}(x)=\Psi(x)\Bigr |_{\stackrel{x_i\,\mapsto x_i+\dl\,(i=i_1,\ldots,i_r)}{x_j\,\mapsto x_j-\dl\,(j=j_1,\ldots,j_r)}}$. By using Jacobi's identity, one can easily see that each of the functions $K_n(\eta;x)$ satisf\/ies the relation
\begin{gather*}
K_{n+1}(\eta;x)
K_{n-1}^{(1234)}\left(\eta;x_0-\dl,x_i-\tfrac{\dl}{2}\right) \\
\qquad{}=K_n^{(24)}\left(\eta;x_0-\tfrac{\dl}{2},x_i-\tfrac{\dl}{4}\right)
K_n^{(13)}\left(\eta;x_0-\tfrac{\dl}{2},x_i-\tfrac{\dl}{4}\right)\\
\qquad{}-K_n^{(14)}\left(\eta;x_0-\tfrac{\dl}{2},x_i-\tfrac{\dl}{4}\right)
K_n^{(23)}\left(\eta;x_0-\tfrac{\dl}{2},x_i-\tfrac{\dl}{4}\right),
\end{gather*}
where $K_n^{(i_1\ldots i_r)}(\eta;x)=K_n(\eta;x)\,|_{\,x_i\mapsto x_i+\dl\,(i=i_1,\ldots,i_r)}$.

\begin{definition}\label{tau_n}
For each $n\in\BZ_{\ge 0}$, we def\/ine the twelve functions $\tau_n(\eta;x)$ by $\tau_n(\eta;x)=\Upsilon_n(\eta;x) K_n(\eta;x)$. The normalization factor $\Upsilon_n(\eta;x)=\Upsilon^{(a;\pm)}_n(x)$ is given by
\begin{gather*}
\Upsilon^{(a;\pm)}_n(x)
 =\dfrac{1}{c_n(x)}
F_{\mp}\left(x_0+\tfrac{1-n}{2}\dl\right)
\disp\prod_{i,j\in C;\,i<j}
F_{\pm\ka^{(a)}_{ij}}\left(x_i+x_j+\tfrac{1-n}{2}\dl\right)\\
\qquad{} \times\CA_{\mp}\left(x_0+\tfrac{1-n}{2}\dl\right)
\disp\prod_{i,j\in C_a;\,i<j}
 \CA_{\pm}\left(x_i+x_j+\tfrac{n+1}{2}\dl\right)
\disp\prod_{i\in C_a}
 \CA_{\mp}\left(x_i+x_a+\tfrac{1-n}{2}\dl\right)\\
 \qquad{}\times
\disp\prod_{i\in C_a}
\CB_{\pm}(x_0+x_i+x_a-\omega-n\dl)
\CB_{\pm}(x_a-x_i+(1-n)\dl)
\CC_{\pm}(x_0+2x_a-\omega-2n\dl),
\end{gather*}
where the functions $F_{\pm}(x)$, $\CA_{\pm}(x)$, $\CB_{\pm}(x)$ and $\CC_{\pm}(x)$ are introduced in Section~\ref{M0}. The factor $c_n(x)$ is def\/ined by
\begin{gather*}
c_n(x)=(-1)^{(\omega+1)\binom{n}{2}}\epsilon_+^{4\binom{n}{2}}
\prod_{r=1}^{n-1}
[x_1-x_2+I_r\dl][x_3-x_4+I_r\dl]\\
\phantom{c_n(x)=}{}\times
\prod_{r=1}^{n-1}
\left[x_1+x_2+\left(r-\tfrac{n+1}{2}\right)\dl\right]^r\left[x_3+x_4+\left(r-\tfrac{n+1}{2}\right)\dl\right]^r,
\end{gather*}
where $I_r\,(r=1,2,\ldots)$ is the subset of $\BZ$ given by $I_r=\{-r+1,-r+3,\ldots,r-3,r-1\}$ and $[x+I_r\dl]=\prod\limits_{k\in I_r}[x+k\dl]$.
\end{definition}

\begin{proposition}
We have the following bilinear relation
\begin{gather}
[x_1-x_2][x_3-x_4]
\tau_{n+1}(\eta;x)
\tau_{n-1}^{(1234)}\left(\eta;x_0-\dl,x_i-\tfrac{\dl}{2}\right)\nonumber\\
=\left[x_1+x_4-\tfrac{n}{2}\dl\right]\left[x_2+x_3-\tfrac{n}{2}\dl\right]
\tau_n^{(24)}\left(\eta;x_0-\tfrac{\dl}{2},x_i-\tfrac{\dl}{4}\right)
\tau_n^{(13)}\left(\eta;x_0-\tfrac{\dl}{2},x_i-\tfrac{\dl}{4}\right)\nonumber\\
{}
-\left[x_2+x_4-\tfrac{n}{2}\dl\right]\left[x_1+x_3-\tfrac{n}{2}\dl\right]
\tau_n^{(14)}\!\left(\eta;x_0-\tfrac{\dl}{2},x_i-\tfrac{\dl}{4}\right)
\tau_n^{(23)}\!\left(\eta;x_0-\tfrac{\dl}{2},x_i-\tfrac{\dl}{4}\right).\!\!\!
  \label{B_tau}
\end{gather}
\end{proposition}

This proposition is easily verif\/ied by noticing that the normalization factor $\Upsilon_n(\eta;x)$ satisf\/ies the relation
\begin{gather*}
[x_1-x_2][x_3-x_4]
\Upsilon_{n+1}(\eta;x)
\Upsilon_{n-1}^{(1234)}\left(\eta;x_0-\dl,x_i-\tfrac{\dl}{2}\right)\\
=\left[x_1+x_4-\tfrac{n}{2}\dl\right]\left[x_2+x_3-\tfrac{n}{2}\dl\right]
\Upsilon_n^{(24)}\left(\eta;x_0-\tfrac{\dl}{2},x_i-\tfrac{\dl}{4}\right)
\Upsilon_n^{(13)}\left(\eta;x_0-\tfrac{\dl}{2},x_i-\tfrac{\dl}{4}\right)\\
{}
=\left[x_2+x_4-\tfrac{n}{2}\dl\right]\left[x_1+x_3-\tfrac{n}{2}\dl\right]
\Upsilon_n^{(14)}\left(\eta;x_0-\tfrac{\dl}{2},x_i-\tfrac{\dl}{4}\right)
\Upsilon_n^{(23)}\left(\eta;x_0-\tfrac{\dl}{2},x_i-\tfrac{\dl}{4}\right).
\end{gather*}

\begin{definition}
For each $\Lam_n\in M_n\,(n\in\BZ)$, we def\/ine the twelve functions $\tau_{\Lam_n}(\eta;x)$ by
\begin{gather}
\tau_{\Lam_n}(\eta;x)=\tau_n\big(\eta;x+l^{(n)} \dl\big),\nonumber\\
l^{(n)}_0=\br{v_0,\Lam_n}+\tfrac{1-n}{2},\qquad
l^{(n)}_i=\br{v_i,\Lam_n}+\tfrac{1-n}{4}\qquad(i\in C),
 \label{def_tau_n}
\end{gather}
where the vectors $v_i$ are def\/ined by $v_0=c-\e_{78}$ and $v_i=\hf\e_{ii9}$ $(i\in C)$ that correspond to the variables $x_i$.
\end{definition}

We show below that the functions $\tau_{\Lam_n}(\eta;x)$ are precisely the hypergeometric $\tau$-func\-tions on~$M_n$. As a preparation, let us def\/ine the action of $\wt{W}\big(D_6^{(1)}\big)$ 
 on the label set $S=\{(a,\epsilon)\,|\,a\in C,\epsilon=\pm 1\}$.

\begin{definition}
We def\/ine the action of $\wt{W}(D_6^{(1)})$ on the label $\eta=(a,\epsilon)\in S$ as follows:
\begin{enumerate}\itemsep=0pt
\item The label is invariant under the action of any translation.

\item The action of any permutation $\sig\in\DS_6$ is def\/ined by $\sig : (a;\pm)\mapsto (\sig(a);\pm)$.

\item Take two mutually distinct indices $i,j\in C$. If $a\notin\{i,j\}$, then $s_{ij7}:(a;\pm)\mapsto (a;\pm)$. Otherwise, we have $s_{ij7}:(a;\pm)\mapsto (b;\mp)$, where $b$ is an index such that $\{a,b\}=\{i,j\}$.

\item The action of the central element $w_c$ is def\/ined by $w_c : (a;\pm)\mapsto (a;\mp)$.
\end{enumerate}
\end{definition}

\begin{theorem}\label{Main}
\indent
\begin{enumerate}\itemsep=0pt
\item[{\rm 1.}] For each $\eta\in S$, the family of functions $\{\tau_{\Lam}(\eta;x)\}_{\Lam\in M^{E_7}}$ satisfies all the bilinear equations for the $q$-Painlev\'e system of type $E_7^{(1)}$ under the conditions \eqref{Rc_tau}  and \eqref{Rc_para}.

\item[{\rm 2.}] For each $n\in\BZ$, the action of $\wt{W}\big(D_6^{(1)}\big)$ on the set of functions $\{\tau_{\Lam_n}(\eta;x)\,|\,\eta\in S,\,\Lam_n\in M_n\}$ is described by $\tau_{w.\Lam_n}(w(\eta);x)=\tau_{\Lam_n}(\eta;w(x))$ for any $w\in\wt{W}\big(D_6^{(1)}\big)$.
\end{enumerate}
\end{theorem}

Let us verify the f\/irst statement. We consider the bilinear equations
\begin{gather}
[\ep_{12}][\ep_{34}]
\tau_{L_{m,1}+\e_8}
\tau_{L_{m,-1}+2\e_0-\e_1-\e_2-\e_3-\e_4-\e_8}\nonumber\\
\qquad{}=[\ep_{148}-m\dl][\ep_{238}-m\dl]
\tau_{L_{m,0}+\e_0-\e_2-\e_4}\tau_{L_{m,0}+\e_0-\e_1-\e_3}\nonumber\\
\qquad{}-[\ep_{248}-m\dl][\ep_{138}-m\dl]
\tau_{L_{m,0}+\e_0-\e_1-\e_4}\tau_{L_{m,0}+\e_0-\e_2-\e_3}
   \label{B_even}
\end{gather}
and
\begin{gather}
[\ep_{12}][\ep_{34}]
\tau_{L_{m,2}+c+\e_{89}+\e_7}
\tau_{L_{m,0}+c+2\e_0-\e_1-\e_2-\e_3-\e_4-\e_8-\e_9+\e_7}\nonumber\\
\qquad{}=[\ep_{148}-m\dl][\ep_{238}-m\dl]
\tau_{L_{m,1}+c+\e_{249}+\e_7}\tau_{L_{m,1}+c+\e_{139}+\e_7}\nonumber\\
\qquad{}-[\ep_{248}-m\dl][\ep_{138}-m\dl]
\tau_{L_{m,1}+c+\e_{149}+\e_7}\tau_{L_{m,1}+c+\e_{239}+\e_7},
  \label{B_odd}
\end{gather}
where $L_{m,n}=m(m+n)c+m\e_{89}$ $(m\in\BZ)$, which are of type (B)${}'_{2m}$ and (B)${}'_{2m+1}$, respectively. Substituting~(\ref{def_tau_n}), we see that these bilinear equations are satisf\/ied thanks to~(\ref{B_tau}).

In order to verify the second statement of Theorem~\ref{Main}, we use the following lemma.

\begin{lemma}\label{ladder}
Suppose that the functions $\tau_{\Lam_{n-1}}(\eta;x)$ and $\tau_{\Lam_n}(\eta;x)$ satisfy all the bilinear equations of type ${\rm (D)}'_n$ and ${\rm (C)}'_{n-1}$, and that we have the relations $\tau_{w.\Lam_{n-1}}(w(\eta);x)=\tau_{\Lam_{n-1}}(\eta;w(x))$ and $\tau_{w.\Lam_n}(w(\eta);x)=\tau_{\Lam_n}(\eta;w(x))$ for any $w\in W(D_6)$. Then the function $\tau_{\Lam_{n+1}}(\eta;x)$ determined by the bilinear equation \eqref{B_even} or \eqref{B_odd} also satisfies $\tau_{w.\Lam_{n+1}}(w(\eta);x)=\tau_{\Lam_{n+1}}(\eta;w(x))$ for any $w\in W(D_6)$.
\end{lemma}

\begin{proof}
From the assumption, we have the bilinear relations
\begin{gather}
[x_1-x_2][x_3-x_4]\tau_n^{(12)}(\eta;x)\tau_n^{(34)}(\eta;x)
+(1,2,3)\mbox{-cyclic}=0,\label{bi_D_n}
\\
[x_3-x_4] \left[x_1+x_5-\tfrac{n{-}1}{2}\dl\right]
\tau_{n-1}^{(1234)} \left(\eta;x_0-\tfrac{\dl}{2},x_i-\tfrac{\dl}{4}\right)
\tau_n^{(25)}(\eta;x)
+(3,4,5)\mbox{-cyclic}=0,\nonumber\\
[x_3-x_4] \left[x_2+x_5-\tfrac{n{-}1}{2}\dl\right]
\tau_{n-1}^{(1234)} \left(\eta;x_0-\tfrac{\dl}{2},x_i-\tfrac{\dl}{4}\right)
\tau_n^{(15)}(\eta;x)
+(3,4,5)\mbox{-cyclic}=0,
   \label{bi_C_n-1}
\end{gather}
and the relations $\tau_{n-1}(w(\eta);x)=\tau_{n-1}(\eta;w(x))$ and $\tau_n(w(\eta);x)=\tau_n(\eta;w(x))$ for any $w\in W(D_6)=\br{s_{12},s_{23},s_{34},s_{45},s_{56},s_{127}}$. What we have to do is to show that the function $\tau_{n+1}(\eta;x)$ determined by the recurrence relation (\ref{B_tau}) also satisf\/ies
\begin{equation}
\tau_{n+1}(w(\eta);x)=\tau_{n+1}(\eta;w(x))\label{cons_tau_n+1}
\end{equation}
for any $w\in W(D_6)$. It is obvious that we have (\ref{cons_tau_n+1}) for $w=s_{12},s_{34},s_{56}$ and $s_{127}$ under the assumption. Then, it is suf\/f\/icient to verify (\ref{cons_tau_n+1}) for $w=s_{23}$ and $s_{45}$. Replacing $x$ by $\wt{x}=s_{23}(x)$ in the recurrence relation (\ref{B_tau}), we get
\begin{gather*}
[x_1-x_3][x_2-x_4]
\tau_{n+1}(\eta;\wt{x})
\tau_{n-1}^{(1234)}\left(\wt{\eta};x_0-\dl,x_i-\tfrac{\dl}{2}\right)\\
\qquad{}\!=\left[x_1+x_4-\tfrac{n}{2}\dl\right]\left[x_2+x_3-\tfrac{n}{2}\dl\right]
\tau_n^{(34)}\left(\wt{\eta};x_0-\tfrac{\dl}{2},x_i-\tfrac{\dl}{4}\right)
\tau_n^{(12)}\left(\wt{\eta};x_0-\tfrac{\dl}{2},x_i-\tfrac{\dl}{4}\right)\\
\qquad{}\!
-\left[x_3+x_4-\tfrac{n}{2}\dl\right]\left[x_1+x_2-\tfrac{n}{2}\dl\right]
\tau_n^{(14)}\left(\wt{\eta};x_0-\tfrac{\dl}{2},x_i-\tfrac{\dl}{4}\right)
\tau_n^{(23)}\left(\wt{\eta};x_0-\tfrac{\dl}{2},x_i-\tfrac{\dl}{4}\right),
\end{gather*}
where $\wt{\eta}=s_{23}(\eta)$. Then, the bilinear equation (\ref{bi_D_n}) yields $\tau_{n+1}(\wt{\eta};x)=\tau_{n+1}(\eta;\wt{x})$. Similarly, replacing $x$ by $\wt{x}=s_{45}(x)$ in the recurrence relation (\ref{B_tau}), we get
\begin{gather*}
[x_1-x_2][x_3-x_5]
\tau_{n+1}(\eta;\wt{x})
\tau_{n-1}^{(1235)}\left(\wt{\eta};x_0-\dl,x_i-\tfrac{\dl}{2}\right)\\
=\left[x_1+x_5-\tfrac{n}{2}\dl\right]\left[x_2+x_3-\tfrac{n}{2}\dl\right]
\tau_n^{(25)}\left(\wt{\eta};x_0-\tfrac{\dl}{2},x_i-\tfrac{\dl}{4}\right)
\tau_n^{(13)}\left(\wt{\eta};x_0-\tfrac{\dl}{2},x_i-\tfrac{\dl}{4}\right)\\
{}
-\left[x_2+x_5-\tfrac{n}{2}\dl\right]\left[x_1+x_3-\tfrac{n}{2}\dl\right]
\tau_n^{(15)}\left(\wt{\eta};x_0-\tfrac{\dl}{2},x_i-\tfrac{\dl}{4}\right)
\tau_n^{(23)}\left(\wt{\eta};x_0-\tfrac{\dl}{2},x_i-\tfrac{\dl}{4}\right),
\end{gather*}
where $\wt{\eta}=s_{45}(\eta)$. From the bilinear relations (\ref{bi_C_n-1}), we get $\tau_{n+1}(\wt{\eta};x)=\tau_{n+1}(\eta;\wt{x})$.
\end{proof}

We already have $\tau_{w.\Lam_0}(w(\eta);x)=\tau_{\Lam_0}(\eta;w(x))$ and $\tau_{w.\Lam_1}(w(\eta);x)=\tau_{\Lam_1}(\eta;w(x))$ for any $w\in W(D_6)$ from Theorems~\ref{act_on_tau0} and~\ref{act_on_tau1}. Also, these functions satisfy all the bilinear equations of type (C)${}'_0$ and (D)${}'_1$ from Theorem~\ref{tau1_bi}. Then we have $\tau_{w.\Lam_2}(w(\eta);x)=\tau_{\Lam_2}(\eta;w(x))$ for any $w\in\wt{W}\big(D_6^{(1)}\big)$ from Lemma~\ref{ladder}. Applying Propositions~\ref{consistency_1}, \ref{consistency_2} and \ref{consistency_3} repeatedly, we can verify the second statement of Theorem~\ref{Main}.

With respect to the system of $q$-dif\/ference equations~(\ref{eq:E7}), one can get the explicit expression for the hypergeometric solutions in terms of the functions $\tau_n(\eta;x)$ introduced in Def\/inition~\ref{tau_n}. When the label $\eta\in S$ is f\/ixed, one can express $b_i$ $(i=1,2,\ldots,8)$ and $t$ in terms of the parameters of the $q$-hypergeometric function ${}_8W_7$. For instance, in the case of $\eta=(6;+)\in S$, we have the following.

\begin{corollary}\label{sol_diff_eq}
Define the functions $f_n(x)$ and $g_n(x)$ by
\[
f_n(x)=\left(\dfrac{qa_0a_4a_5}{a_1a_2a_3}\right)^{1/4}t^{1/2}
\dfrac{N_{f,n}(x)}{D_{f,n}(x)},\qquad
g_n(x)=\left(\dfrac{a_1a_2a_3}{qa_0a_4a_5}\right)^{1/4}t^{1/2}
\dfrac{N_{g,n}(x)}{D_{g,n}(x)}
\]
with
\begin{gather*}
N_{f,n}(x)
 =(a_1/a_3)^{1/4}
\tau_n^{[1]}\left(x_0-\tfrac{3}{2}\dl,x_i+\tfrac{\dl}{4}\right)
\tau_n^{(12)}\left(x_0+\tfrac{3}{2}\dl,x_i-\tfrac{\dl}{4}\right)\\
\phantom{N_{f,n}(x)=}{} -(a_3/a_1)^{1/4}
\tau_n^{[3]}\left(x_0-\tfrac{3}{2}\dl,x_i+\tfrac{\dl}{4}\right)
\tau_n^{(23)}\left(x_0+\tfrac{3}{2}\dl,x_i-\tfrac{\dl}{4}\right),\\
D_{f,n}(x)
=(a_1/a_3)^{1/4}
\tau_n^{[3]}\left(x_0-\tfrac{3}{2}\dl,x_i+\tfrac{\dl}{4}\right)
\tau_n^{(23)}\left(x_0+\tfrac{3}{2}\dl,x_i-\tfrac{\dl}{4}\right)\\
\phantom{D_{f,n}(x)=}{} -(a_3/a_1)^{1/4}
\tau_n^{[1]}\left(x_0-\tfrac{3}{2}\dl,x_i+\tfrac{\dl}{4}\right)
\tau_n^{(12)}\left(x_0+\tfrac{3}{2}\dl,x_i-\tfrac{\dl}{4}\right),\\
N_{g,n}(x)
=(a_2/a_3)^{1/4}
\tau_n^{[3]}\left(x_0-\tfrac{3}{2}\dl,x_i+\tfrac{\dl}{4}\right)
\tau_n^{(13)}\left(x_0+\tfrac{3}{2}\dl,x_i-\tfrac{\dl}{4}\right)\\
\phantom{N_{g,n}(x)=}{}-(a_3/a_2)^{1/4}
\tau_n^{[2]}\left(x_0-\tfrac{3}{2}\dl,x_i+\tfrac{\dl}{4}\right)
\tau_n^{(12)}\left(x_0+\tfrac{3}{2}\dl,x_i-\tfrac{\dl}{4}\right),\\
D_{g,n}(x)
=(a_2/a_3)^{1/4}
\tau_n^{[2]}\left(x_0-\tfrac{3}{2}\dl,x_i+\tfrac{\dl}{4}\right)
\tau_n^{(12)}\left(x_0+\tfrac{3}{2}\dl,x_i-\tfrac{\dl}{4}\right)\\
\phantom{D_{g,n}(x)=}{} -(a_3/a_2)^{1/4}
\tau_n^{[3]}\left(x_0-\tfrac{3}{2}\dl,x_i+\tfrac{\dl}{4}\right)
\tau_n^{(13)}\left(x_0+\tfrac{3}{2}\dl,x_i-\tfrac{\dl}{4}\right),
\end{gather*}
where $a_0,a_1,\ldots,a_5$ are the parameters of the hypergeometric function ${}_8W_7(a_0;a_1,\ldots,a_5;q,z)$ defined by \eqref{ep<->a}, $\tau_n^{(i_1i_2)}(x)=\tau_n(x)|_{x_i\mapsto x_i+\dl\,(i=i_1,i_2)}$ and $\tau_n^{[i]}(x)=\tau_n(x)|_{x_i\mapsto x_i-\dl}$. Let $b_{i,n}$ $(i=1,2,\ldots,8)$ be the parameters defined by
\begin{alignat*}{3}
&b_{1,n}=-q^{1/4}a_3^{-3/4}(a_0a_4a_5)^{1/4},&&
b_{2,n}=-q^{1/4}a_4^{-3/4}(a_0a_3a_5)^{1/4},&\\
&b_{3,n}=-q^{1/4}a_5^{-3/4}(a_0a_3a_4)^{1/4},&&
b_{4,n}=-q^{1/4}a_0^{-3/4}(a_3a_4a_5)^{1/4},&\\
&b_{5,n}=-q^{1/4}(a_1a_2)^{-1/2}(a_0a_3a_4a_5)^{1/4},&&
b_{6,n}=-q^{5/4}a_0^{5/4}(a_1a_2)^{-1/2}(a_3a_4a_5)^{-3/4},&\\
&b_{7,n}=-q^{n/2-3/4}a_0^{-3/4}(a_1a_2)^{1/2}(a_3a_4a_5)^{1/4},\quad \ &&
b_{8,n}=-q^{-n/2-3/4}a_0^{-3/4}(a_1a_2)^{1/2}(a_3a_4a_5)^{1/4}&
\end{alignat*}
and $t=(a_1/a_2)^{1/2}$. Then, $f=f_n(x)$ and $g=g_n(x)$ with $b_i=b_{i,n}$ give rise to a solution of the system of $q$-difference equations~\eqref{eq:E7}.
\end{corollary}

\appendix
\section[The $q$-Painlev\'e system of type $E_7^{(1)}$]{The $\boldsymbol{q}$-Painlev\'e system of type $\boldsymbol{E_7^{(1)}}$}

\subsection[Point configurations and Cremona transformations]{Point conf\/igurations and Cremona transformations}

As mentioned in Section~\ref{qP_E7}, the $q$-Painlev\'e system of type $E_7^{(1)}$ is a discrete dynamical system def\/ined on a family of rational surfaces parameterized by nine-point conf\/igurations on $\BP^2$ such that six points among them are on a conic $C$ and other three are on a line $L$~\cite{Sakai2}. In this section, we set $p_1,p_2,p_4,p_5,p_6,p_7\in C$ and $p_3,p_8,p_9\in L$ so that the standard Cremona transformation with respect to
$(p_1,p_2,p_3)$ is well-def\/ined as a birational action on the conf\/iguration space. One can parameterize the conf\/iguration space by~\cite{Sakai2,KMNOY2}
\[
X=\left[
\begin{array}{cccccccccc}
1&1&-u_3& 1 & 1& 1 & 1& -u_8&-u_9& x_1\\
u_1&u_2&0&u_4&u_5&u_6&u_7&0&0& x_2\\
u_1^2&u_2^2&1&u_4^2&u_5^2&u_6^2&u_7^2&1&1& x_3
\end{array}
\right],
\]
where $u_1,u_2,\ldots,u_9$ are parameters satisfying $u_1u_2\cdots u_9=q^{-1}$ $(q\in\BC^*)$ and the tenth column denotes the coordinates of a general point on $\BP^2$. The symmetric group $\DS_6^C\times\DS_3^L$ with $\DS_6^C=\br{s_{12},s_{24},s_{45},s_{56},s_{67}}$ and $\DS_3^L=\br{s_{38},s_{89}}$ naturally acts on the space as $\sig(u_j)=u_{\sig(j)}$ for any $\sig\in\DS_6^C\times\DS_3^L$.

Let us normalize $X$ by an action of $GL_3(\BC)$ as
\[
Y=\left[
\begin{array}{ccccccc}
1&0&0&u_{14}&\ldots&u_{19}& y_1\\
0&1&0&u_{24}&\ldots&u_{29}& y_2\\
0&0&1&u_{34}&\ldots&u_{39}& y_3
\end{array}
\right].
\]
The coordinates $u_{ij}$ can be expressed by
\begin{equation}
u_{ij}=\left\{
\begin{array}{ll}
y_i^C(u_j)\quad& (j=1,2,4,5,6,7),\\[1mm]
y_i^L(u_j)\quad& (j=3,8,9),
\end{array}
\right. \label{u_of_y}
\end{equation}
where $y_i^C(u)$ and $y_i^L(u)$ are def\/ined by
\begin{gather*}
y_1^C(u)=\dfrac{(u_2-u)(1-u_2u_3u)}
               {(u_2-u_1)(1-u_1u_2u_3)},\qquad
y_2^C(u)=\dfrac{(u_1-u)(1-u_1u_3u)}
               {(u_1-u_2)(1-u_1u_2u_3)},\\
y_3^C(u)=\dfrac{(u_1-u)(u_2-u)}
               {(1-u_1u_2u_3)}
\end{gather*}
and
\begin{gather*}
y_1^L(u)=\dfrac{u_2(u_3-u)}
               {(u_2-u_1)(1-u_1u_2u_3)},\qquad
y_2^L(u)=\dfrac{u_1(u_3-u)}
               {(u_1-u_2)(1-u_1u_2u_3)},\\
y_3^L(u)=\dfrac{1-u_1u_2u}
               {1-u_1u_2u_3},
\end{gather*}
respectively. We further normalize $X$ (or $Y$) by an action of $PGL_3(\BC)$ as
\[
Z=
\left[
\begin{array}{cccccccc}
1 & 0 & 0 & 1 & v_{15} & \ldots & v_{19} & z_1\\
0 & 1 & 0 & 1 & v_{25} & \ldots & v_{29} & z_2\\
0 & 0 & 1 & 1 & 1 & \ldots & 1 & 1
\end{array}
\right],
\]
where the coordinates $v_{ij}$ and $z_i$ are expressed by
\begin{equation}
v_{ij}=\dfrac{u_{34}\,u_{ij}}{u_{i4}\,u_{3j}},\qquad
z_i=\dfrac{u_{34}\,y_i}{u_{i4}\,y_3}\qquad
(i=1,2;\,j=5,6,7,8,9).\label{v_of_u}
\end{equation}
The action of the standard Cremona transformation with respect to $(p_1,p_2,p_3)$, denoted by~$s_{123}$, on these variables is given by $s_{123}(v_{ij})=1/v_{ij}$ and $s_{123}(z_i)=1/z_i$. This transformation together with the symmetric group $\DS_6^C\times\DS_3^L$ generates the af\/f\/ine Weyl group $W(E_7^{(1)})=\br{s_{12},s_{24},s_{45},s_{56},s_{67},s_{38},s_{89},s_{123}}$ associated with the Dynkin diagram
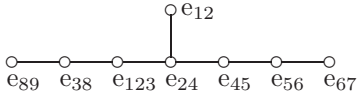
\[
\begin{picture}(140,25)(0,-5)
\put(0,0){\circle{4}}\put(2,0){\line(1,0){16}}
\put(20,0){\circle{4}}\put(22,0){\line(1,0){16}}
\put(40,0){\circle{4}}\put(42,0){\line(1,0){16}}
\put(60,0){\circle{4}}\put(62,0){\line(1,0){16}}
\put(80,0){\circle{4}}\put(82,0){\line(1,0){16}}
\put(100,0){\circle{4}}\put(102,0){\line(1,0){16}}
\put(120,0){\circle{4}}
\put(60,20){\circle{4}}\put(60,2){\line(0,1){16}}
\put(-2,-9){\small$\e_{89}$}
\put(18,-9){\small$\e_{38}$}
\put(38,-9){\small$\e_{123}$}
\put(58,-9){\small$\e_{24}$}
\put(78,-9){\small$\e_{45}$}
\put(98,-9){\small$\e_{56}$}
\put(118,-9){\small$\e_{67}$}
\put(64,18){\small$\e_{12}$}
\end{picture}
\]
In this realization of $W(E_7^{(1)})\subset W(E_8^{(1)})=\br{s_{12},s_{23},\ldots,s_{89},s_{123}}$, the automorphism of the above Dynkin diagram can be expressed by $\pi=s_{124}\,s_{35}\,s_{68}\,s_{79}$. The action of the extended af\/f\/ine Weyl group $\wt{W}(E_7^{(1)})=\br{s_{12},s_{24},s_{45},s_{56},s_{67},s_{38},s_{89},s_{123},\pi}$ on the conf\/iguration space is given by
\begin{alignat}{5}
& s_{12}(z_1)=z_2,\qquad && s_{12}(z_2)=z_1,\qquad && s_{38}(z_1)=\dfrac{z_1-v_{18}}{1-v_{18}},\qquad  &&
s_{38}(z_2)=\dfrac{z_2-v_{28}}{1-v_{28}},& \nonumber\\
& s_{123}(z_1)=\dfrac{1}{z_1},\qquad && s_{123}(z_2)=\dfrac{1}{z_2},\qquad && s_{24}(z_1)=\dfrac{z_2-z_1}{z_2-1},\qquad &&
s_{24}(z_2)=\dfrac{z_2}{z_2-1},&  \label{W_on_z}\\
& s_{45}(z_1)=\dfrac{z_1}{v_{15}},\qquad && s_{45}(z_2)=\dfrac{z_2}{v_{25}} && &\nonumber
\end{alignat}
and
\begin{equation}
\pi(v_{ij})=\dfrac{v_{ik}-v_{i5}}{v_{ik}-1}
\quad (k=8,9,6,7\quad\mbox{for}\quad j=6,7,8,9),\qquad
\pi(z_i)=\dfrac{z_i-v_{i5}}{z_i-1}
 \label{pi_on_z}
\end{equation}
for $i=1,2$. From this representation, we obtain a family of functional equations
\begin{equation}
w(v_{ij})=S_{ij}^w(v),\qquad w(z_i)=R_i^w(v;z)
\label{fe_v}
\end{equation}
for each $w\in\wt{W}(E_7^{(1)})$, where $S_{ij}^w(v)$ and $R_i^w(v;z)$ are some rational functions.

Let us introduce the variables $c_i=e(\ep_i)$ $(i=0,1,\ldots,9)$, where $\ep_i$ are the coordinate functions introduced in Section~\ref{qP_E7}, and suppose that the parameters $u_1,\ldots,u_9$ are expressed by $u_j=c_j/c_0^r$ $(j=1,2,4,5,6,7)$ and $u_j=c_j/c_0^s$ $(j=3,8,9)$ with $2r+s=1$. Then, $y_i^C(u)$ and $y_i^L(u)$ are expressed by
\begin{gather}
y_1^C(t/c_0^r)=\dfrac{(1-t/c_2)(1-c_2c_3t/c_0)}
                     {(1-c_1/c_2)(1-c_1c_2c_3/c_0)},\qquad
y_2^C(t/c_0^r)=\dfrac{(1-t/c_1)(1-c_1c_3t/c_0)}
                     {(1-c_2/c_1)(1-c_1c_2c_3/c_0)},\nonumber\\
y_3^C(t/c_0^r)=\dfrac{c_1c_2(1-t/c_1)(1-t/c_2)}
                     {c_0^{2r}(1-c_1c_2c_3/c_0)}
  \label{cano_sol_C}
\end{gather}
and
\begin{gather}
y_1^L(t/c_0^s)=\dfrac{c_3(1-t/c_3)}
                     {c_0^s(1-c_1/c_2)(1-c_1c_2c_3/c_0)},\qquad
y_2^L(t/c_0^s)=\dfrac{c_3(1-t/c_3)}
                     {c_0^s(1-c_2/c_1)(1-c_1c_2c_3/c_0)},\nonumber\\
y_3^L(t/c_0^s)=\dfrac{1-c_1c_2t/c_0}
                     {1-c_1c_2c_3/c_0},
  \label{cano_sol_L}
\end{gather}
respectively. These expressions with (\ref{u_of_y}) and (\ref{v_of_u})
give us
\begin{gather}
v_{1j}=\dfrac{(u_1-u_4)(1-u_2u_3u_j)}
             {(1-u_2u_3u_4)(u_1-u_j)}
=\dfrac{[\ep_{14}][\ep_{23j}]}
       {[\ep_{234}][\ep_{1j}]},\nonumber\\
v_{2j}=\dfrac{(u_2-u_4)(1-u_1u_3u_j)}
             {(1-u_1u_3u_4)(u_2-u_j)}
=\dfrac{[\ep_{24}][\ep_{13j}]}
       {[\ep_{134}][\ep_{2j}]}
 \label{v_ep_1}
\end{gather}
for $j=5,6,7$ and
\begin{gather}
v_{1j}=\dfrac{u_2(u_1-u_4)(u_3-u_j)}
             {(1-u_2u_3u_4)(1-u_1u_2u_j)}
=\dfrac{[\ep_{14}][\ep_{3j}]}
       {[\ep_{234}][\ep_{12j}]},\nonumber\\
v_{2j}=\dfrac{u_1(u_2-u_4)(u_3-u_j)}
             {(1-u_1u_3u_4)(1-u_1u_2u_j)}
=\dfrac{[\ep_{24}][\ep_{3j}]}
       {[\ep_{134}][\ep_{12j}]}
 \label{v_ep_2}
\end{gather}
for $j=8,9$. We f\/ind that these expressions satisfy $w(v_{ij})(\ep)=v_{ij}(w(\ep))$ for any $w\in\wt{W}\big(E_7^{(1)}\big)$, namely (\ref{v_ep_1}) and (\ref{v_ep_2}) give a solution to the f\/irst equation of~(\ref{fe_v}). Also, we see that the functions
\begin{equation}
z_1=\dfrac{[\ep_{14}]}{[\ep_{234}]}
    \dfrac{[\ep_{123}+\ep_1-t]}{[\ep_1-t]},\qquad
z_2=\dfrac{[\ep_{24}]}{[\ep_{134}]}
    \dfrac{[\ep_{123}+\ep_2-t]}{[\ep_2-t]}
\label{cano_C}
\end{equation}
and
\begin{equation}
z_1=\dfrac{[\ep_{14}]}{[\ep_{234}]}
    \dfrac{[\ep_3-t]}{[\ep_{123}+\ep_3-t]},\qquad
z_2=\dfrac{[\ep_{24}]}{[\ep_{134}]}
    \dfrac{[\ep_3-t]}{[\ep_{123}+\ep_3-t]}
\label{cano_L}
\end{equation}
satisfy $z_i(w(\ep);t)=R_i^w(\ep;z(\ep;t))$ for any $w\in\wt{W}\big(E_7^{(1)}\big)$. This means that each of~(\ref{cano_C}) and~(\ref{cano_L}) provides a one-parameter family of solutions to the second equation of~(\ref{fe_v}). These solutions will be called the canonical solution, which correspond to the vertical solution in the context of the dif\/ferential Painlev\'e equations.

\subsection[The lattice $\tau$-functions and the bilinear equations]{The lattice $\boldsymbol{\tau}$-functions and the bilinear equations}

Here, we introduce a framework of the lattice $\tau$-functions and show that the action of the extended af\/f\/ine Weyl group $\wt{W}(E_7^{(1)})$ is transformed into the bilinear equations for the lattice $\tau$-functions.

Recall that the lattice $\tau$-functions for the discrete Painlev\'e system of type $E_8^{(1)}$ are indexed by $\Lam\in M=W\big(E_8^{(1)}\big)\,.\,\e_1=\{\,\Lam\in\CL\,|\,\br{\Lam,\Lam}=1,\,\br{c,\Lam}=-1\,\}$~\cite{KMNOY1}. Let us decompose the central element $c=3\e_0-\e_1-\e_2-\cdots-\e_9$ into two irreducible components by~\cite{Sakai2}
\[
c={\cal D}_C+{\cal D}_L,\qquad
\left\{
\begin{array}{l}
{\cal D}_C=2\e_0-\e_1-\e_2-\e_4-\e_5-\e_6-\e_7,\\
{\cal D}_L=\e_0-\e_3-\e_8-\e_9
\end{array}\right.
\]
corresponding to the conic $C$ and the line $L$. Then, we have two $W(E_7^{(1)})$-orbits
\begin{gather*}
M^C=\{\Lam\in M\,|\,
    \br{{\cal D}_C,\Lam}=-1,\,\br{{\cal D}_L,\Lam}=0\,\}
   =W\big(E_7^{(1)}\big)\,.\,\e_1,\\[1mm]
M^L=\{\Lam\in M\,|\,
    \br{{\cal D}_C,\Lam}=0,\,\br{{\cal D}_L,\Lam}=-1\,\}
   =W\big(E_7^{(1)}\big)\,.\,\e_3,
\end{gather*}
which are transformed by the action of the Dynkin diagram automorphism $\pi\in\wt{W}\big(E_7^{(1)}\big)$ to each other. Hereafter, we consider the lattice $\tau$-functions $\tau_{\Lam}$ for $\Lam\in M^{E_7}=M^C\coprod M^L=\wt{W}\big(E_7^{(1)}\big) \e_1$, on which the action of $w\in\wt{W}\big(E_7^{(1)}\big)$ is def\/ined by $w(\tau_{\Lam})=\tau_{w.\Lam}$.

Suppose that the variables $y_i$ are expressed by
\[
y_1=\dfrac{\tau_{\e_2}\tau_{\e_3}\tau_{\e_0-\e_2-\e_3}}{N_1},\qquad
y_2=\dfrac{\tau_{\e_1}\tau_{\e_3}\tau_{\e_0-\e_1-\e_3}}{N_2},\qquad
y_3=\dfrac{\tau_{\e_1}\tau_{\e_2}\tau_{\e_0-\e_1-\e_2}}{N_3},
\]
where the normalization factors $N_1$, $N_2$ and $N_3$ are certain functions of $\ep_i$ $(i=0,1,\ldots,9)$. Denote the $\tau$-functions for the canonical solution on the conic $C$ by $\tau_{\Lam}|_C$ $(\Lam\in M^{E_7})$;
\begin{gather*}
y_1^C=\dfrac{\tau_{\e_2}|_C\,\tau_{\e_3}|_C\,\tau_{\e_0-\e_2-\e_3}|_C}{N_1},
\qquad
y_2^C=\dfrac{\tau_{\e_1}|_C\,\tau_{\e_3}|_C\,\tau_{\e_0-\e_1-\e_3}|_C}{N_2},
\\
y_3^C=\dfrac{\tau_{\e_1}|_C\,\tau_{\e_2}|_C\,\tau_{\e_0-\e_1-\e_2}|_C}{N_3}.
\end{gather*}
Comparing this expression with (\ref{cano_sol_C}) and (\ref{cano_sol_L}), one can assume that the $\tau$-functions for the canonical solutions on $C$ and $L$ are expressed by
\[
\tau_{\e_j}|_C=
\left\{
\begin{array}{ll}
\left(1-e(t-\ep_j)\right)e(\aa\ep_j)&(j=1,2,4,5,6,7),\\[1mm]
e(\beta\ep_j)&(j=3,8,9),
\end{array}
\right.
\]
and
\[
\tau_{\e_j}|_L=
\left\{
\begin{array}{ll}
e(\beta\ep_j)&(j=1,2,4,5,6,7),\\[1mm]
\left(1-e(t-\ep_j)\right)e(\aa\ep_j)&(j=3,8,9),
\end{array}
\right.
\]
respectively, so that the canonical solutions $y_i^C$ and $y_i^L$ are transformed by the action of the Dynkin diagram automorphism $\pi$ to each other. These requirements lead us to $r=1/4$, $s=1/2$ and $\beta=\aa-1/2$, and we get $N_1=-c_0^{\aa-1/2}c_1[\ep_{12}][\ep_{123}]$, $N_2=c_0^{\aa-1/2}c_2[\ep_{12}][\ep_{123}]$ and $N_3=c_0^{\aa-1/2}c_3^{1/2}[\ep_{123}]$.

Let us introduce the variables $f_i\,(i=1,2,3)$ by $f_1=\dfrac{\tau_{\e_0-\e_2-\e_3}}{\tau_{\e_1}}$, $f_2=\dfrac{\tau_{\e_0-\e_1-\e_3}}{\tau_{\e_2}}$ and $f_3=\dfrac{\tau_{\e_0-\e_1-\e_2}}{\tau_{\e_3}}$. Then, the inhomogeneous coordinates $z_1$ and $z_2$ are expressed by
\begin{equation}
z_1=\dfrac{[\ep_{14}]}{[\ep_{234}]}\dfrac{f_1}{f_3},\qquad
z_2=\dfrac{[\ep_{24}]}{[\ep_{134}]}\dfrac{f_2}{f_3}.\label{z_f}
\end{equation}
From (\ref{W_on_z}), (\ref{pi_on_z}) and (\ref{z_f}), one thus obtain a realization of the extended af\/f\/ine Weyl group $\wt{W}\big(E_7^{(1)}\big)$ as a group of birational transformations.

\begin{theorem}
The action of $\wt{W}\big(E_7^{(1)}\big)$ on the variables $(f_1,f_2,f_3)$ and $(\tau_{\e_1},\ldots,\tau_{\e_9})$ is given~by
\begin{gather*}
\sig(\tau_{\e_i})=\tau_{\e_{\sig(i)}}\qquad
(\sig\in\DS_6^C\times\DS_3^L),\\
s_{123}(\tau_{\e_1})=\tau_{\e_1}f_1,\qquad
s_{123}(\tau_{\e_2})=\tau_{\e_2}f_2,\qquad
s_{123}(\tau_{\e_3})=\tau_{\e_3}f_3,\\
s_{12}(f_1)=f_2,\qquad s_{12}(f_2)=f_1,\\
s_{38}(f_1)=
\dfrac{\tau_{\e_3}}{\tau_{\e_8}}
\dfrac{[\ep_{128}]f_1-[\ep_{38}]f_3}{[\ep_{123}]},\qquad
s_{38}(f_2)=
\dfrac{\tau_{\e_3}}{\tau_{\e_8}}
\dfrac{[\ep_{128}]f_2-[\ep_{38}]f_3}{[\ep_{123}]},\qquad
s_{38}(f_3)=\dfrac{\tau_{\e_3}}{\tau_{\e_8}}f_3,\\
s_{123}(f_1)=\dfrac{1}{f_1},\qquad
s_{123}(f_2)=\dfrac{1}{f_2},\qquad
s_{123}(f_3)=\dfrac{1}{f_3},\\
s_{24}(f_1)=
\dfrac{\tau_{\e_2}}{\tau_{\e_4}}
\dfrac{[\ep_{14}][\ep_{134}]f_1-[\ep_{24}][\ep_{234}]f_2}
      {[\ep_{12}][\ep_{123}]},\qquad
s_{24}(f_2)=\dfrac{\tau_{\e_2}}{\tau_{\e_4}}f_2,\\
s_{24}(f_3)=
\dfrac{\tau_{\e_2}}{\tau_{\e_4}}
\dfrac{[\ep_{134}]f_3-[\ep_{24}]f_2}{[\ep_{123}]},
\end{gather*}
and
\begin{gather*}
\pi(\tau_{\e_1})=
\dfrac{\tau_{\e_1}\tau_{\e_3}}{\tau_{\e_4}}
\dfrac{-[\ep_{14}]f_1+[\ep_{234}]f_3}{[\ep_{123}]},\qquad
\pi(\tau_{\e_2})=
\dfrac{\tau_{\e_2}\tau_{\e_3}}{\tau_{\e_4}}
\dfrac{-[\ep_{24}]f_2+[\ep_{134}]f_3}{[\ep_{123}]},\\
\pi(\tau_{\e_3})=\tau_{\e_5},\qquad\pi(\tau_{\e_4})=\tau_{\e_3}f_3,\qquad
\pi(\tau_{\e_5})=\tau_{\e_3},\\
\pi(\tau_{\e_6})=\tau_{\e_8},\qquad\pi(\tau_{\e_7})=\tau_{\e_9},\qquad
\pi(\tau_{\e_8})=\tau_{\e_6},\qquad\pi(\tau_{\e_9})=\tau_{\e_7},\\
\pi(f_1)=\dfrac{\tau_{\e_4}}{\tau_{\e_5}}
\dfrac{[\ep_{15}]f_1-[\ep_{235}]f_3}
      {[\ep_{14}]f_1-[\ep_{234}]f_3},\qquad
\pi(f_2)=\dfrac{\tau_{\e_4}}{\tau_{\e_5}}
\dfrac{[\ep_{25}]f_2-[\ep_{135}]f_3}
      {[\ep_{24}]f_2-[\ep_{134}]f_3},\qquad
\pi(f_3)=\dfrac{\tau_{\e_4}}{\tau_{\e_5}}.
\end{gather*}
These give rise to a representation of $\wt{W}\big(E_7^{(1)}\big)$.
\end{theorem}

From this theorem, we immediately obtain the bilinear equations (\ref{bi_1}) and (\ref{bi_2}) for mutually distinct indices $i,j,k,l\in\{1,2,4,5,6,7\}$ and $r,s\in\{3,8,9\}$.

\section{Another representation}
In this section we again set $C=\{1,2,3,4,5,6\}$ and $L=\{7,8,9\}$. The lattice $\tau$-functions~$\tau_{\Lam}$ $(\Lam\in M^{E_7})$ for the $q$-Painlev\'e system of type $E_7^{(1)}$ satisfy the following bilinear equations
\begin{gather}
[\ep_{rs}]\tau_{\e_j}\tau_{\e_0-\e_i-\e_j}
=[\ep_{ijs}]\tau_{\e_r}\tau_{\e_0-\e_i-\e_r}
-[\ep_{ijr}]\tau_{\e_s}\tau_{\e_0-\e_i-\e_s}, \nonumber\\
[\ep_{jk}]\tau_{\e_r}\tau_{\e_0-\e_i-\e_r}
=[\ep_{ikr}]\tau_{\e_j}\tau_{\e_0-\e_i-\e_j}
-[\ep_{ijr}]\tau_{\e_k}\tau_{\e_0-\e_i-\e_k},
 \label{bi_1'}
\end{gather}
and
\begin{gather*}
[\ep_{ij}][\ep_{ijr}]\tau_{\e_k}\tau_{\e_0-\e_k-\e_r}
+(i,j,k)\mbox{-cyclic}=0,\\
[\ep_{ij}][\ep_{kl}]\tau_{\e_0-\e_i-\e_j}\tau_{\e_0-\e_k-\e_l}
+(i,j,k)\mbox{-cyclic}=0,
\end{gather*}
where $i,j,k,l\in C$ and $r,s\in L$.

As discussed in~\cite{KMNOY1,KMNOY4}, when $\Lam=d\e_0-\nu_1\e_1-\cdots-\nu_9\e_9$, the $\tau$-function $\tau_{\Lam}$ is characterized by a homogeneous polynomial of degree $d$ in the homogeneous coordinates of $\BP^2$ which has a~zero of multiplicity $\ge\nu_j$ at $p_j$ for each $j=1,\ldots,9$. From this geometric consideration, we f\/ind that we have the following bilinear equation
\begin{equation}
\tau_{\e_i}\tau_{\e_0-\e_i-\e_9}-
\tau_{\e_j}\tau_{\e_0-\e_j-\e_9}+
[\ep_{ij}][\ep_{ij9}]\,d_L\tau_{\e_7}\tau_{\e_8}=0
\label{bi_3'}
\end{equation}
for $i,j\in C$, which associates with the line passing through the point $p_9$. The factor $d_L$ corresponds to the irreducible component of the anti-canonical devisor ${\cal D}_L=\e_0-\e_7-\e_8-\e_9$, and is invariant under the action of $W\big(E_7^{(1)}\big)$. The action of the Dynkin diagram automorphism $\pi$ on this bilinear equation gives us the second equation in (\ref{bi_3}) with $d_C=\pi(d_L)$.

One can get another representation $\wt{W}\big(E_7^{(1)}\big)$ for the $\tau$-variables, by using the bilinear equations (\ref{bi_1'}) and (\ref{bi_3'}).

\begin{theorem}
Let us introduce the variables $\sig$ and $\wt{\sig}$ by
\begin{gather*}
\sig=d_r
\dfrac{e( \frac{1}{4}\ep_{23})\tau_{\e_3}\tau_{\e_0-\e_1-\e_3}
      -e(-\frac{1}{4}\ep_{23})\tau_{\e_2}\tau_{\e_0-\e_1-\e_2}}
      {[\ep_{23}]},\\
\wt{\sig}=d_l
\dfrac{e( \frac{1}{4}\ep_{23})\tau_{\e_2}\tau_{\e_0-\e_1-\e_2}
      -e(-\frac{1}{4}\ep_{23})\tau_{\e_3}\tau_{\e_0-\e_1-\e_3}}
      {[\ep_{23}]},
\end{gather*}
where the factors $d_l$ and $d_r$ are given by $d_l=e\left(\frac{1}{16} \aa_l-\frac{1}{16} \aa_r\right)$ and $d_r=d_l^{-1}$ with $\aa_l=3\ep_{127}+2\ep_{78}+\ep_{89}$ and $\aa_r=3\ep_{34}+2\ep_{45}+\ep_{56}$. Then, the action of $\wt{W}\big(E_7^{(1)}\big)$ on the variables $\tau_{\e_3}$, $\tau_{\e_4}$, $\tau_{\e_5}$, $\tau_{\e_6}$, $\tau_{\e_7}$, $\tau_{\e_8}$, $\tau_{\e_9}$, $\tau_{\e_0-\e_1-\e_2}$, $\sig$ and $\wt{\sig}$ is described as follows:
\begin{gather}
s_{89}: \ \tau_{\e_8}\lra\tau_{\e_9},\qquad
s_{78}: \ \tau_{\e_7}\lra\tau_{\e_8}, \qquad
s_{127}: \ \tau_{\e_7}\lra\tau_{\e_0-\e_1-\e_2},\nonumber\\
s_{34}: \ \tau_{\e_3}\lra\tau_{\e_4},\qquad
s_{45}: \ \tau_{\e_4}\lra\tau_{\e_5},\qquad
s_{56}: \ \tau_{\e_5}\lra\tau_{\e_6},\nonumber
\\
s_{23}(\tau_{\e_3})=
\dfrac{e(-\frac{1}{4}\ep_{23})\,d_r^{-1}\sig
      +e( \frac{1}{4}\ep_{23})\,d_l^{-1}\wt{\sig}}
      {\tau_{\e_0-\e_1-\e_2}},\nonumber\\
s_{23}(\tau_{\e_0-\e_1-\e_2})=
\dfrac{e(\frac{1}{4}\ep_{23})\,d_r^{-1}\sig
      +e(-\frac{1}{4}\ep_{23})\,d_l^{-1}\wt{\sig}}
      {\tau_{\e_3}},
 \label{s23}
\\
s_{12}(\sig)=
\dfrac{e(-\frac{1}{4}\ep_{12})
         \tau_{\e_7}\tau_{\e_8}\tau_{\e_9}\tau_{\e_0-\e_1-\e_2}+
       e( \frac{1}{4}\ep_{12})
         \tau_{\e_3}\tau_{\e_4}\tau_{\e_5}\tau_{\e_6}}{\wt{\sig}},\nonumber\\
s_{12}(\wt{\sig})=
\dfrac{e(\frac{1}{4}\ep_{12})
         \tau_{\e_7}\tau_{\e_8}\tau_{\e_9}\tau_{\e_0-\e_1-\e_2}+
       e(-\frac{1}{4}\ep_{12})
         \tau_{\e_3}\tau_{\e_4}\tau_{\e_5}\tau_{\e_6}}{\sig},
 \label{s12}
\\
\pi: \ \tau_{\e_3}\lra\tau_{\e_0-\e_1-\e_2},\qquad
         \tau_{\e_4}\lra\tau_{\e_7},\qquad
         \tau_{\e_5}\lra\tau_{\e_8},\qquad
         \tau_{\e_6}\lra\tau_{\e_9},\qquad\sig\lra\wt{\sig}.\nonumber
\end{gather}
These also give rise to another representation of $\wt{W}\big(E_7^{(1)}\big)$.
\end{theorem}

\begin{proof}
We immediately get (\ref{s23}) from the def\/inition of $\sig$ and $\wt{\sig}$. It is easy to see that $\sig$ and $\wt{\sig}$ are invariant under the action of $s_{89}$, $s_{78}$, $s_{127}$, $s_{23}$, $s_{34}$, $s_{45}$ and $s_{56}$. The bilinear equations~(\ref{bi_1'}) and (\ref{bi_3'}) yield
\begin{gather*}
\left[
e\left( \tfrac{1}{4}\ep_{23}\right)\tau_{\e_2}\tau_{\e_0-\e_1-\e_2}-
e\left(-\tfrac{1}{4}\ep_{23}\right)\tau_{\e_3}\tau_{\e_0-\e_1-\e_3}
\right]\\
\qquad{}\times
\left[
e\left( \tfrac{1}{4}\ep_{13}\right)\tau_{\e_3}\tau_{\e_0-\e_2-\e_3}-
e\left(-\tfrac{1}{4}\ep_{13}\right)\tau_{\e_1}\tau_{\e_0-\e_1-\e_2}
\right]\\
\qquad{}=[\ep_{13}][\ep_{23}]
\left[
e\left(-\tfrac{1}{4}\ep_{12}\right)\tau_{\e_7}\tau_{\e_8}\tau_{\e_9}\tau_{\e_0-\e_1-\e_2}+
e\left( \tfrac{1}{4}\ep_{12}\right)\tau_{\e_3}\tau_{\e_4}\tau_{\e_5}\tau_{\e_6}
\right],
\end{gather*}
from which we get the f\/irst equation of (\ref{s12}). Since we see that $\pi :\sig\lra\wt{\sig}$ by the def\/inition, we immediately get the second equation of (\ref{s12}).
\end{proof}

Note that this representation coincides with that constructed by Tsuda~\cite{Tsuda}. The above theorem gives us the following proposition.

\begin{proposition}
Define the variables $f$ and $g$ by
\[
f=\dfrac{\wt{\sig}}{\sig}
\dfrac{e( \frac{1}{4}\ep_{12})
          \tau_{\e_7}\tau_{\e_8}\tau_{\e_9}\tau_{\e_0-\e_1-\e_2}
      +e(-\frac{1}{4}\ep_{12})
          \tau_{\e_3}\tau_{\e_4}\tau_{\e_5}\tau_{\e_6}}
      {e(-\frac{1}{4}\ep_{12})
          \tau_{\e_7}\tau_{\e_8}\tau_{\e_9}\tau_{\e_0-\e_1-\e_2}
      +e( \frac{1}{4}\ep_{12})
          \tau_{\e_3}\tau_{\e_4}\tau_{\e_5}\tau_{\e_6}},\qquad
g=\dfrac{\wt{\sig}}{\sig}.
\]
Then, the action of $\wt{W}(E_7^{(1)})$ on these variables is described by
\[
s_{12}: \ f\lra g,\qquad
\pi: \ f\mapsto\dfrac{1}{f},\qquad g\mapsto\dfrac{1}{g},\qquad
s_{23}: \ f\mapsto h,
\]
where $h$ is a rational function determined by
\begin{equation}
\dfrac{h+d_l^2e(\hf\ep_{12})}{h+d_l^2e(-\hf\ep_{12})}=
\dfrac{f+d_l^2e(\hf\ep_{13})}{f+d_l^2e(-\hf\ep_{13})}
\dfrac{g+d_l^2e(-\hf\ep_{23})}{g+d_l^2e(\hf\ep_{23})}.\label{fgh}
\end{equation}
\end{proposition}

Note that the variable $f$ can be expressed by
\begin{equation}
f=d_l^2\,\dfrac
{e( \frac{1}{4}\ep_{13})\tau_{\e_1}\tau_{\e_0-\e_1-\e_2}
-e(-\frac{1}{4}\ep_{13})\tau_{\e_3}\tau_{\e_0-\e_2-\e_3}}
{e( \frac{1}{4}\ep_{13})\tau_{\e_3}\tau_{\e_0-\e_2-\e_3}
-e(-\frac{1}{4}\ep_{13})\tau_{\e_1}\tau_{\e_0-\e_1-\e_2}},
\label{f_tau}
\end{equation}
and $h=s_{13}(g)$.

\section[A derivation of the difference equations]{A derivation of the dif\/ference equations} \label{derivation}

By writing down the action of the translation operator $T_{21}\in W\big(E_7^{(1)}\big)$ on the variables $f$ and $g$, we will get the system of $q$-dif\/ference equations~(\ref{eq:E7}). Hereafter, we denote the time evolution of a variable $x$ by $\ol{x}=T_{21}(x)$ and $\ul{x}=T_{21}^{-1}(x)$. Let us introduce the transformation $\mu$ by $\mu=s_{12}s_{23}s_{147}s_{158}s_{169}$. It is easy to see that $T_{21}=\mu^2$ and $\mu(g)=f$. We also introduce the auxiliary variables $k=s_{247}(h)$ and $l=s_{258}(k)$. Note that we have $\ol{g}=s_{269}(l)$.

\begin{lemma}
We have
\begin{gather*}
\dfrac{f+d_l^2e(-\hf\ep_{13})}
      {f+d_l^2e( \hf\ep_{13})}
=\dfrac{f/g-e(-\hf\ep_{12})}
       {f/g+e(-\hf\ep_{12})}
\dfrac{f/h-e(-\hf\ep_{23})}
      {f/h-e( \hf\ep_{23})},\\
\dfrac{f+d_l^2\ka_{47}e(-\hf\ep_{247})}
      {f+d_l^2\ka_{47}e( \hf\ep_{247})}
=\dfrac{f/h-e( \hf\ep_{23})}
       {f/h-e(-\hf\ep_{23})}
\dfrac{f/k-e(-\hf\ep_{347})}
      {f/k-e( \hf\ep_{347})},\\
\dfrac{f+d_l^2\ka_{58}e(-\hf\ep_{258})}
      {f+d_l^2\ka_{58}e( \hf\ep_{258})}
=\dfrac{f/k-e( \hf\ep_{347})}
       {f/k-e(-\hf\ep_{347})}
\dfrac{f/l-e(-\hf\ep_{347}-\hf\ep_{258})}
      {f/l-e( \hf\ep_{347}+\hf\ep_{258})},\\
\dfrac{f+d_l^2\ka_{69}e(-\hf\ep_{269})}
      {f+d_l^2\ka_{69}e( \hf\ep_{269})}
=\dfrac{f/l-e( \hf\ep_{347}+\hf\ep_{258})}
       {f/l-e(-\hf\ep_{347}-\hf\ep_{258})}
\dfrac{f/\ol{g}-e(-\hf(\ep_{12}+\dl))}
      {f/\ol{g}-e( \hf(\ep_{12}+\dl))}.
\end{gather*}
\end{lemma}

\begin{proof}
The f\/irst equation is reduced to (\ref{fgh}). The other expressions can be rewritten as
\begin{gather}
\dfrac{f+d_l^2\ka_{47}e(-\hf\ep_{247})}
      {f+d_l^2\ka_{47}e( \hf\ep_{247})}
\dfrac{h+d_l^2\ka_{47}e( \hf\ep_{347})}
      {h+d_l^2\ka_{47}e(-\hf\ep_{347})}
\dfrac{k+d_l^2\ka_{47}e(-\hf\ep_{23})}
      {k+d_l^2\ka_{47}e( \hf\ep_{23})}=1,\nonumber\\
\dfrac{f+d_l^2\ka_{58}e(-\hf\ep_{258})}
      {f+d_l^2\ka_{58}e( \hf\ep_{258})}
\dfrac{k+d_l^2\ka_{58}e( \hf\ep_{347}+\hf\ep_{258})}
      {k+d_l^2\ka_{58}e(-\hf\ep_{347}-\hf\ep_{258})}
\dfrac{l+d_l^2\ka_{58}e(-\hf\ep_{347})}
      {l+d_l^2\ka_{58}e( \hf\ep_{347})}=1,\nonumber\\
\dfrac{f+d_l^2\ka_{69}e(-\hf\ep_{269})}
      {f+d_l^2\ka_{69}e( \hf\ep_{269})}
\dfrac{l+d_l^2\ka_{69}e( \hf(\ep_{12}+\dl))}
      {l+d_l^2\ka_{69}e(-\hf(\ep_{12}+\dl))}
\dfrac{\ol{g}+d_l^2\ka_{69}e(-\hf\ep_{347}-\hf\ep_{258})}
      {\ol{g}+d_l^2\ka_{69}e( \hf\ep_{347}+\hf\ep_{258})}=1,
 \label{eqs}
\end{gather}
where $\ka_{47}=e(\hf\ep_{34}-\hf\ep_{127})$, $\ka_{58}=e(\hf\ep_{35}-\hf\ep_{128})$ and $\ka_{69}=e(\hf\ep_{36}-\hf\ep_{129})$. From the expressions (\ref{f_tau}), we get
\[
\dfrac{f+d_l^2\ka_{47}e(-\hf\ep_{247})}
      {f+d_l^2\ka_{47}e( \hf\ep_{247})}
=e\left(-\tfrac 12 \ep_{247}\right)
\dfrac{\tau_{\e_7}\tau_{\e_0-\e_2-\e_7}}
      {\tau_{\e_4}\tau_{\e_0-\e_2-\e_4}}.
\]
By applying $s_{13}s_{12}$ and $s_{247}$ successively, we also get
\begin{gather*}
\dfrac{h+d_l^2\ka_{47}e( \hf\ep_{347})}
      {h+d_l^2\ka_{47}e(-\hf\ep_{347})}
=e\left(\tfrac 12 \ep_{347}\right)
\dfrac{\tau_{\e_4}\tau_{\e_0-\e_3-\e_4}}
      {\tau_{\e_7}\tau_{\e_0-\e_3-\e_7}},\\
\dfrac{k+d_l^2\ka_{47}e(-\hf\ep_{23})}
      {k+d_l^2\ka_{47}e( \hf\ep_{23})}
=e\left(-\tfrac 12 \ep_{23}\right)
\dfrac{\tau_{\e_0-\e_2-\e_4}\tau_{\e_0-\e_3-\e_7}}
      {\tau_{\e_0-\e_2-\e_7}\tau_{\e_0-\e_3-\e_4}},
\end{gather*}
and then the f\/irst equation of (\ref{eqs}). The second and third equations of (\ref{eqs}) can be obtained by a similar way.
\end{proof}

The above lemma immediately gives us
\begin{gather*}
\dfrac{fg-e(\hf\ep_{12})}
      {fg-e(-\hf\ep_{12})}
\dfrac{f\ol{g}-e(\hf(\ep_{12}+\dl))}
      {f\ol{g}-e(-\hf(\ep_{12}+\dl))}\\
\qquad{}=
\dfrac{f+d_l^2e(\hf\ep_{13})}
      {f+d_l^2e(-\hf\ep_{13})}
\dfrac{f+d_l^2\ka_{47}e( \hf\ep_{247})}
      {f+d_l^2\ka_{47}e(-\hf\ep_{247})}
\dfrac{f+d_l^2\ka_{58}e( \hf\ep_{258})}
      {f+d_l^2\ka_{58}e(-\hf\ep_{258})}
\dfrac{f+d_l^2\ka_{69}e( \hf\ep_{269})}
      {f+d_l^2\ka_{69}e(-\hf\ep_{269})},
\end{gather*}
where we replace $g$ with $1/g$. Applying $\mu^{-1}$ to the above equation, we also get
\begin{gather*}
\dfrac{\ul{f}g-e(-\hf(\ep_{12}-\dl))}
      {\ul{f}g-e(\hf(\ep_{12}-\dl))}
\dfrac{fg-e(-\hf\ep_{12})}
      {fg-e(\hf\ep_{12})}\\
\qquad{}=
\dfrac{g+d_r^2e(\hf\ep_{23})}
      {g+d_r^2e(-\hf\ep_{23})}
\dfrac{g+d_r^2\ka_{47}^{-1}e( \hf\ep_{147})}
      {g+d_r^2\ka_{47}^{-1}e(-\hf\ep_{147})}
\dfrac{g+d_r^2\ka_{58}^{-1}e( \hf\ep_{158})}
      {g+d_r^2\ka_{58}^{-1}e(-\hf\ep_{158})}
\dfrac{g+d_r^2\ka_{69}^{-1}e( \hf\ep_{169})}
      {g+d_r^2\ka_{69}^{-1}e(-\hf\ep_{169})}.
\end{gather*}
Let us introduce the parameters $b_i$ $(i=1,2,\ldots,8)$ and the independent variable $t$ by
\begin{alignat*}{3}
& b_1=-q^{1/8}d_l^2e\left(\tfrac 14(\ep_{13}+\ep_{23})\right),&&
b_2=-q^{1/8}d_l^2e\left(\tfrac 14(\ep_{14}+\ep_{24})+\tfrac 12\ep_{34}\right),& \\
&b_3=-q^{1/8}d_l^2e\left(\tfrac 14(\ep_{15}+\ep_{25})+\tfrac 12\ep_{35}\right),&&
b_4=-q^{1/8}d_l^2e\left(\tfrac 14(\ep_{16}+\ep_{26})+\tfrac 12\ep_{36}\right),&\\
& b_5=-q^{1/8}d_l^2e\left(-\tfrac 14(\ep_{13}+\ep_{23})\right),&&
b_6=-q^{1/8}d_l^2e\left(-\tfrac 14(\ep_{137}+\ep_{237})-\tfrac 12\ep_{127}\right),& \\
& b_7=-q^{1/8}d_l^2e\left(-\tfrac 14(\ep_{138}+\ep_{238})-\tfrac 12\ep_{128}\right),\quad \ &&
b_8=-q^{1/8}d_l^2e\left(-\tfrac 14(\ep_{139}+\ep_{239})-\tfrac 12\ep_{129}\right)&
\end{alignat*}
and $t=e(\hf\ep_{12})$, respectively. Replacing the dependent variables $f$ and $g$ with $q^{-1/8}t^{-1/2}f$ and $q^{1/8}t^{-1/2}g$, respectively, we get the system of dif\/ference equations (\ref{eq:E7}).

\subsection*{Acknowledgements}
The author would like to express his sincere thanks to Professors M.~Noumi and Y.~Yamada for valuable discussions and comments. Especially, he owes initial steps of this work, including the formulation by means of the lattice $\tau$-functions and the bilinear equations, to discussions with them. The author would also thank Professors K.~Kajiwara and Y.~Ohta for stimulating discussions.

\pdfbookmark[1]{References}{ref}
\LastPageEnding

\end{document}